%% file: main.tex
\documentclass[preprint]{elsarticle}

\usepackage[utf8]{inputenc}
\usepackage[colorlinks,bookmarksopen,bookmarksnumbered,citecolor=black,urlcolor=black,linkcolor=black]{hyperref}
\usepackage{graphicx}
\usepackage{listings, lstautogobble}
\usepackage{xcolor}
\usepackage[T1]{fontenc}
\usepackage{amssymb}
\usepackage[ruled, linesnumbered]{algorithm2e}
\usepackage{caption}
\usepackage{balance}
\usepackage{enumitem}
\usepackage[normalem]{ulem}
\definecolor{codebackground}{rgb}{1,1,1}
\definecolor{black}{rgb}{0.13, 0.55, 0.13}

\newcommand{\Let}{\textbf{let~}}

\newcommand{\Output}{\textbf{output~}}

\lstdefinelanguage{PSEUDO}{
	morekeywords={let, new, Set, for, all, in, if, Claim, axiomatic, assumed, output, +, else},
	sensitive=true,
	morecomment=[l]{//},
	morecomment=[s]{/*}{*/},
	morestring=[b]",
	showstringspaces=false
}

\lstdefinelanguage{EOL}{
	morekeywords={@template,out,TemplateFactory,Template,delete,import,for,while,in,and,or,self,operation,return,def,var,throw,if,new,else,transaction,abort,
		break,breakall,continue,assert,assertError,not, switch, case, default, println, first, all, select, size, includes},
	sensitive=true,
	morecomment=[l]{//},
	morecomment=[s]{/*}{*/},
	morestring=[b]",
	showstringspaces=false
}

\lstdefinelanguage{Emfatic}{
	morekeywords={class, extends, attr, float},
	sensitive=true,
	morecomment=[l]{//},
	morecomment=[s]{/*}{*/},
	morestring=[b]",
	showstringspaces=false
}

\begin{document}

		\title{ACCESS: Assurance Case Centric Engineering of Safety-critical Systems}
		
		\author[cam,lan]{Ran Wei}
        \ead{rw741@cam.ac.uk; r.wei5@lancaster.ac.uk}
		
		\author[york]{Simon Foster\corref{mycorrespondingauthor}}
		\ead{simon.foster@york.ac.uk}

        \author[york]{Haitao Mei\corref{mycorrespondingauthor}}
		\ead{haitao.mei@icloud.com}
		
		\author[york]{Fang Yan\corref{mycorrespondingauthor}}
		\ead{fang.yan@york.ac.uk}
		
		\author[dut]{Ruizhe Yang\corref{mycorrespondingauthor}}
		\ead{ruizheyang@mail.dlut.edu.cn}
		
		\author[york]{Ibrahim Habli}
        \author[drisq]{Colin O'Halloran}
        \author[drisq]{Nick Tudor}
        \author[york]{Tim Kelly}
        \author[york]{Yakoub Nemouchi}
		
		\address[cam]{Department of Engineering, University of Cambridge, Cambridge, CB3 0FA, UK}
        \address[lan]{School of Computing and Communications, University of Lancaster, Lancaster, LA1 4WA, UK}
        \address[york]{Department of Computer Science, University of York, York, YO10 5GH, UK}
        \address[dut]{School of Computer Science and Technology, Dalian University of Technology, Dalian, 116024, China}
        \address[drisq]{D-RisQ Ltd., Malvern, WR14 3SZ, UK}
		
		\cortext[mycorrespondingauthor]{Corresponding authors}

\begin{abstract}
Assurance cases are used to communicate and assess confidence in critical system properties such as safety and security. Historically, assurance cases have been manually created documents, which are evaluated by system stakeholders through lengthy and complicated processes. In recent years, model-based system assurance approaches have gained popularity to improve the efficiency and quality of system assurance activities. This becomes increasingly important, as systems becomes more complex, it is a challenge to manage their development life-cycles, including coordination of development, verification and validation activities, and change impact analysis in inter-connected system assurance artifacts. Moreover, there is a need for assurance cases that support evolution during the operational life of the system, to enable continuous assurance in the face of an uncertain environment, as Robotics and Autonomous Systems (RAS) are adopted into society. In this paper, we contribute ACCESS - Assurance Case Centric Engineering of Safety-critical Systems, an engineering methodology, together with its tool support, for the development of safety critical systems around evolving model-based assurance cases. We show how model-based system assurance cases can trace to heterogeneous engineering artifacts (e.g. system architectural models, system safety analysis, system behaviour models, etc.), and how formal methods can be integrated during the development process. We demonstrate how assurance cases can be automatically evaluated both at development and runtime. We apply our approach to a case study based on an Autonomous Underwater Vehicle (AUV). 
\end{abstract}

\maketitle
\input{section_1}
\input{section_2}
\input{section_3}
\input{section_4}
\input{section_5}
\input{section_6}

\input{section_7}
\input{section_8}

\section*{Acknowledgement}
Simon Foster's contributions are funded by the EPSRC-UKRI project CyPhyAssure (https://www.cs.york.ac.uk/circus/CyPhyAssure/), grant reference EP/S001190/1. Fang Yan's contributions are funded by the European Union’s Horizon 2020 research and innovation programme under the Marie Skłodowska-Curie grant agreement No 812.788 (MSCA-ETN SAS).
\balance

\bibliographystyle{abbrv}
\bibliography{references}

\end{document}

%% file: section_1.tex
\section{Introduction}
\label{sec:introduction}
Safety-critical systems require justifications that they are acceptably safe to operate in their defined operational contexts.
\textit{Assurance case}s provide an explicit means for arguing, justifying and assessing the confidence in the safety of safety-critical systems. 
The submission of an assurance case is increasingly being required during system certification in many safety-critical industries, such as aviation~\cite{eurocontrol}, nuclear power~\cite{iaea}, transportation~\cite{iso26262, ukrail}, healthcare \cite{habli2018safety} and defence~\cite{ukmod}. 
Prior to certification, an assurance case must be rigorously, and often independently, \textit{evaluated}\footnote{In this work, we use \textit{evaluation} to refer to both validation and verification activities involved in the development and the assessment processes of safety critical systems and their assurance cases.} to ensure that the arguments and evidence for safety is coherent and convincing.

Assurance cases are not self-contained documents. 
They usually depend on a variety of \textit{engineering artifacts} that provide contextual and evidential information, including requirement documents, architecture designs, behaviour models, safety analyses, etc.
These artifacts may originate from diverse languages and tools, and can be used for prototyping, analysis, formal verification, and the derivation of real-world artifacts.
Therefore, assurance case \textit{evaluation} involves the evaluation of the engineering artifacts an assurance case depends on, which is often an informal, manual, and error-prone process.
In an \textit{idealised} development process, an assurance case is the central point of reference for all system stakeholders, to allow effective communication over diverse engineering artifacts.
In addition, changes in engineering artifacts require the assurance case to be re-evaluated \cite{denney2015dynamic}, which can significantly impact development efficiency. This challenge becomes more obvious as systems become more complex.
Hence, there is a need to automate some (if not all) of the system assurance activities to efficiently manage assurance cases and their referenced engineering artifacts.

Over the past few years, system assurance practitioners have begun adopting Model Based Systems Engineering (MBSE). 
MBSE promises the interoperability and management of diverse artifacts/models in an automated manner, which provide the basis for automated, coherent and self-contained assurance cases.
However, existing assurance case notations, such as the Goal Structuring Notation (GSN)~\cite{kelly2004goal} and Claim-Argument-Evidence (CAE)~\cite{bishop2000methodology}, do not have a sufficient model-based foundation to systematically support this kind of integration~\cite{wei2019model}.
Consequently, existing model-based assurance case approaches cannot provide the collective and automated evaluation of an assurance case, together with the engineering artifacts that it may depend on.
The inspection, evaluation, and change management of engineering artifacts still remain manual.

In recent years, new applications for Robotics and Autonomous Systems (RAS) have emerged, which are often safety critical.
RASs are increasingly open (they inter-connect at runtime) and adaptive (they adapt to changing contexts at runtime), that render the current generation of safety assurance approaches insufficient~\cite{trapp2013safety, denney2015dynamic}. 
Specifically, assurance cases for RAS need to be \textit{living} documents that can evolve during the operational life of the system with minimal human intervention. 
As such, it is imperative to shift some system safety assurance activities from development time to runtime~\cite{trapp2013safety, wei2018}.
This is a significant challenge, though, as it requires automation of verification and validation activities that are both crucial parts of the evaluation process.
On the one hand, we must demonstrate that each engineering artifact meets its requirements through verification, and on the other we must ensure that real-world system artifacts exhibit the behaviour predicted by a model through techniques like runtime safety monitoring~\cite{Machin2018SMOF}.

To address the identified challenges, we introduce our model-based, assurance oriented methodology -- Assurance Case Centric Engineering of Safety-critical Systems (ACCESS).
ACCESS is underpinned by a combination of (1) design-time automated assurance-case-and-engineering-artifacts management and evaluation, and (2) runtime assurance case evaluation based on runtime data.
We present a tooling prototype, Assurance Case Management Environment (ACME), which supports the creation and the management of assurance cases based on the Structured Assurance Case Metamodel (SACM) \cite{sacm}, an international standard. 

To demonstrate our approach, we provide a case study on the assurance case for an Autonomous Underwater Vehicle (AUV), including its safety requirements, arguments, and a formal model of the safety controller in the RoboChart language~\cite{Miyazawa2019-RoboChart} with formal verification evidence. 

We discuss how we can apply our approach to develop critical systems around an evolving assurance case.
In addition, we also demonstrate tool support for ACCESS at development time, so that we can: 1) perform a collective evaluation of an assurance case, via model-based traceability, with respect to the engineering artifacts it refers to; 2) automatically invoke evidence from formal methods by using Isabelle/HOL~\cite{Isabelle} as a verification service within an assurance case; 3) automatically generate and machine-check formalisation of an assurance case to verify its logical integrity; and 4) enable automated change impact analysis from engineering artifacts to assurance cases.
We also discuss how we can turn a development time assurance case to a dynamic runtime assurance case and discuss how runtime assurance case evaluation (based on runtime data) can be achieved using our approach. 

The main contributions of our paper are:
\begin{enumerate}
    \item ACCESS -- a critical systems engineering methodology around an evolving assurance case model;
    \item Automated means to evaluate an assurance case with its referenced engineering artifacts at development time and runtime (with a prototypical dynamic assurance case management system to evaluate assurance cases based on runtime data);
    \item Facilities to integrate diverse formal verification results into an assurance case and automatic generation of a formalised assurance case in Isabelle/SACM for analysis using theorem proving;
    \item Automated change impact analysis from engineering artifacts to assurance cases;
    \item The application of all of above to an AUV case study.
\end{enumerate}

The rest of the paper is organised as follows. 
In Section~\ref{sec:preliminaries} we provide some background information on assurance cases, GSN, model-based assurance case and formal methods. 
In Section~\ref{sec:access} we describe the generic ACCESS methodology. 
In Section~\ref{sec:toolsupport} we discuss our tool support to back the ACCESS methodology
In Section~\ref{sec:case} we evaluate the ACCESS methodology in depth with an case study for an Autonomous Underwater Vehicle (AUV).
In Section~\ref{sec:related} we discuss related work and in Section~\ref{sec:conclusion} we conclude this paper with a discussion and point out future research directions.

%% file: section_2.tex
\section{Preliminaries}
\label{sec:preliminaries}
This section provides background information on assurance cases, assurance case notations, model-based system assurance and formal methods, including the terminologies and concepts used in the rest of the paper. 

\subsection{Assurance Cases}
\label{sec:acs}
The practice of safety certification is increasingly \textit{goal-oriented} rather than highly prescriptive~\cite{denney2015dynamic}. 
Goal-oriented certification places greater emphasis on explicitly stating safety claims, and supplying an argument along with supporting evidence to satisfy certification goals that regulators define \cite{mcdermid2001software}. 
Examples may include ``All identified hazards have been mitigated'' or ``Necessary assumptions of the physical environment have been defined''. 
Such arguments and evidence are generally organised in the form of an \textit{assurance case}.
As defined in \cite{kelly1999arguing}, an assurance case is a document that \textit{should communicate a clear, comprehensive and defensible argument that a system is acceptably safe to operate in a particular context}. 

\begin{figure}[h]
    \centering
	\includegraphics[width=1\columnwidth]{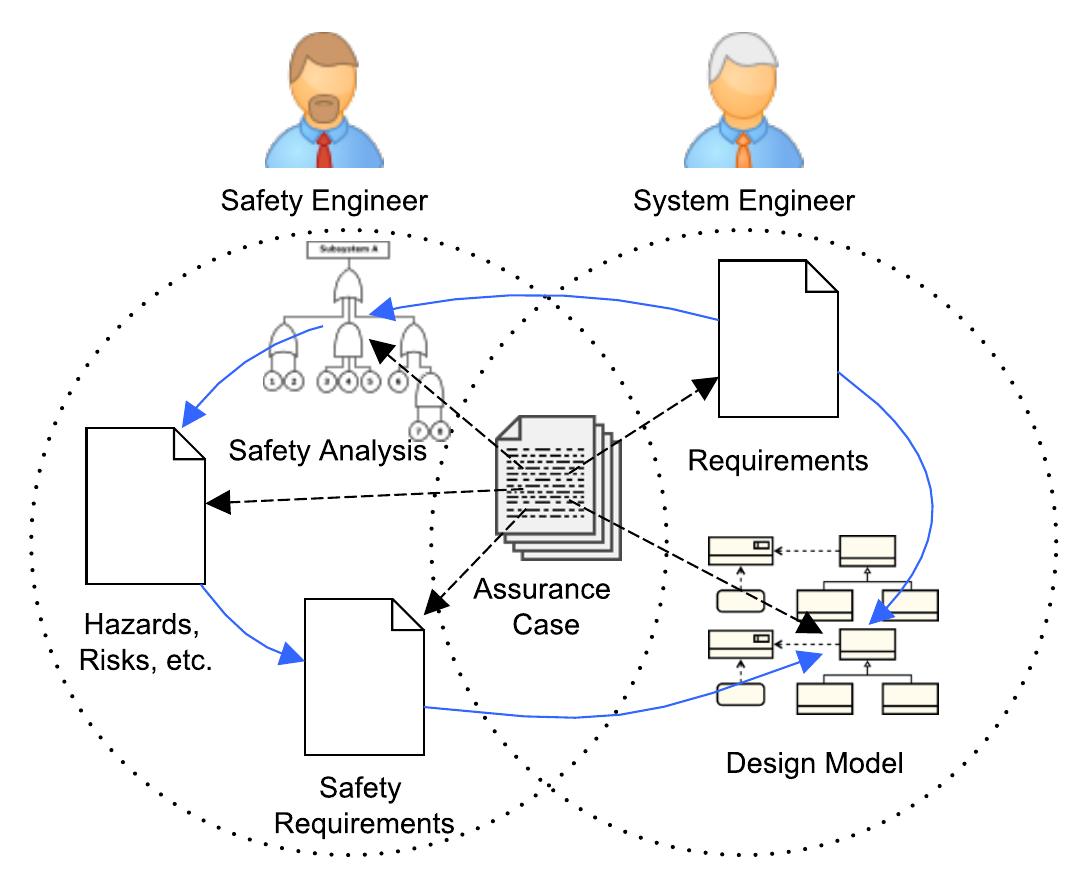}
	\caption{Assurance cases and engineering artifacts.}
	\label{fig:sc_dependencies}
\end{figure}

Conventionally, an assurance case is not a self-contained document. 
Definitions of assurance case~\cite{jsp430, 0055} indicate that an assurance case is a document (either as a logical concept or as a physical artifact) that can refer to, and pull together information regarding system safety (such as system requirements, system architectural design, safety analyses, etc.), to form a safety argument.
The development of an assurance case involves communications among various stakeholders, one typical scenario is illustrated in Figure~\ref{fig:sc_dependencies}. 
When a system concept is formed, system engineers define a set of requirements. 
Based on these requirements, safety engineers may perform safety analyses, from which hazards (and their associated risks) are identified. 
Identified hazards and risks are then used to derive safety goals, from which safety engineers can elicit safety requirements.
Safety requirements are then considered in the system design, which impose constraints (e.g. acceptable failure rates), and mitigation measures (e.g. redundancy and monitoring).
As an assurance case is developed, it may refer to all engineering artifacts above within its argument, for contextual and evidential information.
Therefore, the \textit{evaluation} of an assurance case typically involves the validation and verification of engineering artifacts it refers to.

Assurance cases are subject to evolution, where engineering artifacts and arguments need to be adapted, potentially because of new requirements or upgrades.
Changes in the engineering artifacts may invalidate the assurance case, and so its integrity must be checked through evaluation. 
To evaluate an assurance case, practitioners typically need to trace, navigate to, review, validate and verify the engineering artifacts it depends on~\cite{hawkins2015weaving}. 
This then informs the decision as to whether the system is acceptably safe to deploy and operate in its intended operational context. 
On the other hand, when an engineering artifact is changed during the development process, its impact in the assurance case needs to be identified, sometimes resulting the assurance case to be re-evaluated.
This is observed in~\cite{nair2015evidence}, the authors of which report that evidence completeness and change impact for assurance cases are managed mostly manually using (sometimes even no) traceability information. 
Importantly, their study raises the question of how evolution and changes are identified, assessed and managed at the level of assurance case.

\subsection{Goal Structuring Notation}
\label{sec:gsn}
\begin{figure}[h]
	\centering
	\includegraphics[width=1\linewidth]{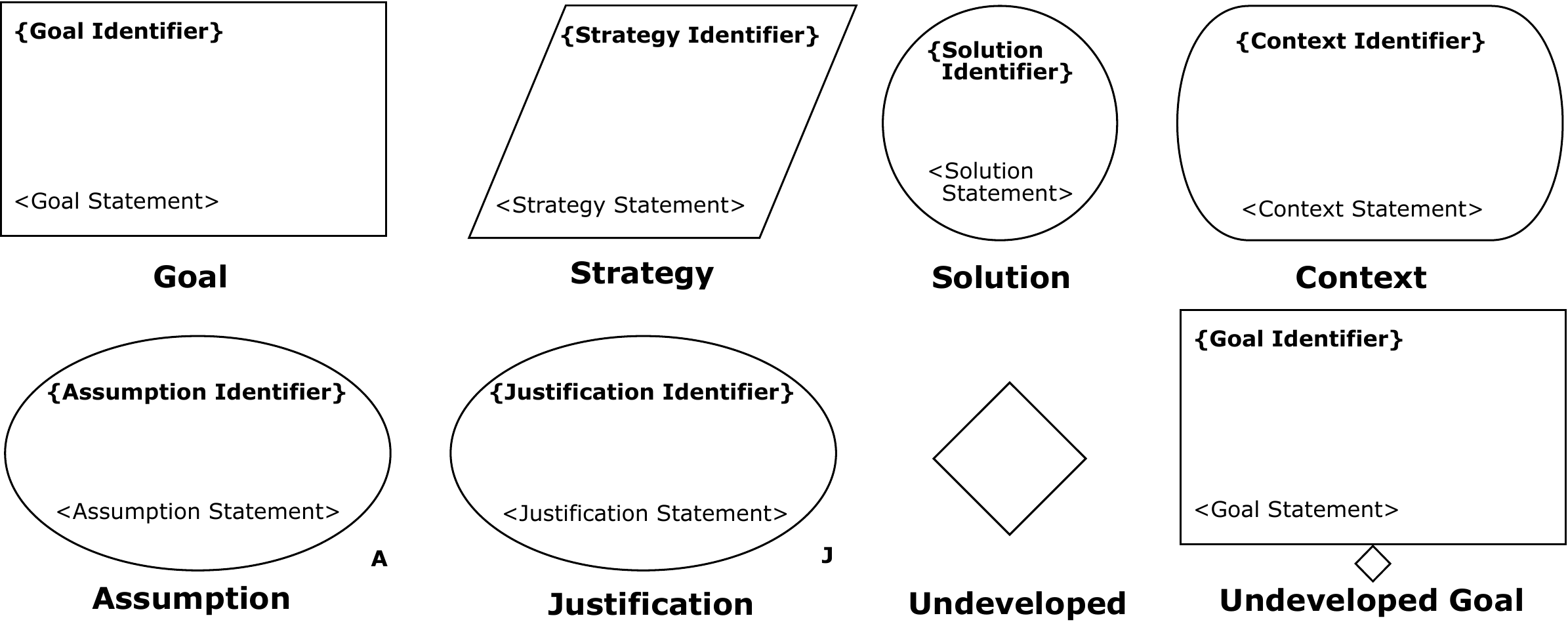}
	\caption{Core GSN elements.}
	\label{fig:gsnCore}
\end{figure}

In the current state of practice, assurance cases are typically communicated using graphical notations, among which the most widely used notation is the Goal Structuring Notation (GSN)~\cite{kelly2004goal}.
GSN is a well established graphical argumentation notation that is widely adopted within safety-critical industries for the presentation of safety arguments within safety cases. 
The core elements of GSN are shown in Figure~\ref{fig:gsnCore}.

A \textit{Goal} represents a safety claim within the argumentation. 
A \textit{Strategy} is used to describe the nature of the inference that exists between a goal and its supporting goal(s). 
A \textit{Solution} represents a reference to an evidence item or multiple evidence items. 
A \textit{Context} represents a contextual artefact, which can be a statement, or a reference to contextual information. 
An \textit{Assumption} represents an assumed statement made within the argumentation. 
A \textit{Justification} represents a statement of rationale. 
An element can be \textit{Undeveloped}, which means that a line of argument has not been developed yet (meaning it being abstract and needs to be instantiated). 
The \textit{Undeveloped} notation can apply to \textit{Goal}s and \textit{Strategies}. 
The \textit{Undeveloped Goal} in Figure~\ref{fig:gsnCore} is an example. 

\begin{figure}[ht!]
	\centering
 	\includegraphics[width=0.7\linewidth]{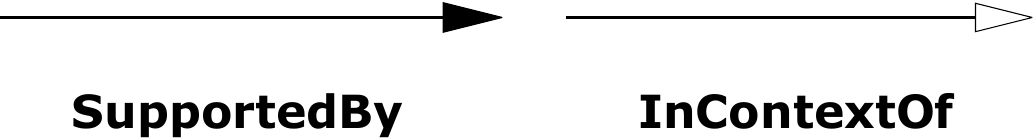}
	\caption{GSN connectors.}
	\label{fig:gsnEdges}
\end{figure}

Core elements of GSN are connected with two types of connectors, as shown in Figure~\ref{fig:gsnEdges}. 
The \textit{SupportedBy} connector allows inferential or evidential relationships to be documented. 
The \textit{InContextOf} relates contextual elements (i.e. \textit{Context}, \textit{Assumption} and \textit{Justification}) to \textit{Goal}s and \textit{Strategies}.

When elements of GSN are linked together in a network, they are often referred to as a \textit{goal structure}. 
The purpose of a goal structure is to show how \textit{Goal}s are successively broken down into sub-\textit{Goal}s until a point is reached where \textit{Goal}s can be supported by direct reference to available evidence (\textit{Solution}s).
An example of a goal structure is shown in Figure~\ref{fig:auv_GSN_module}.

\begin{figure}[ht!]
	\centering
	\includegraphics[width=1\linewidth]{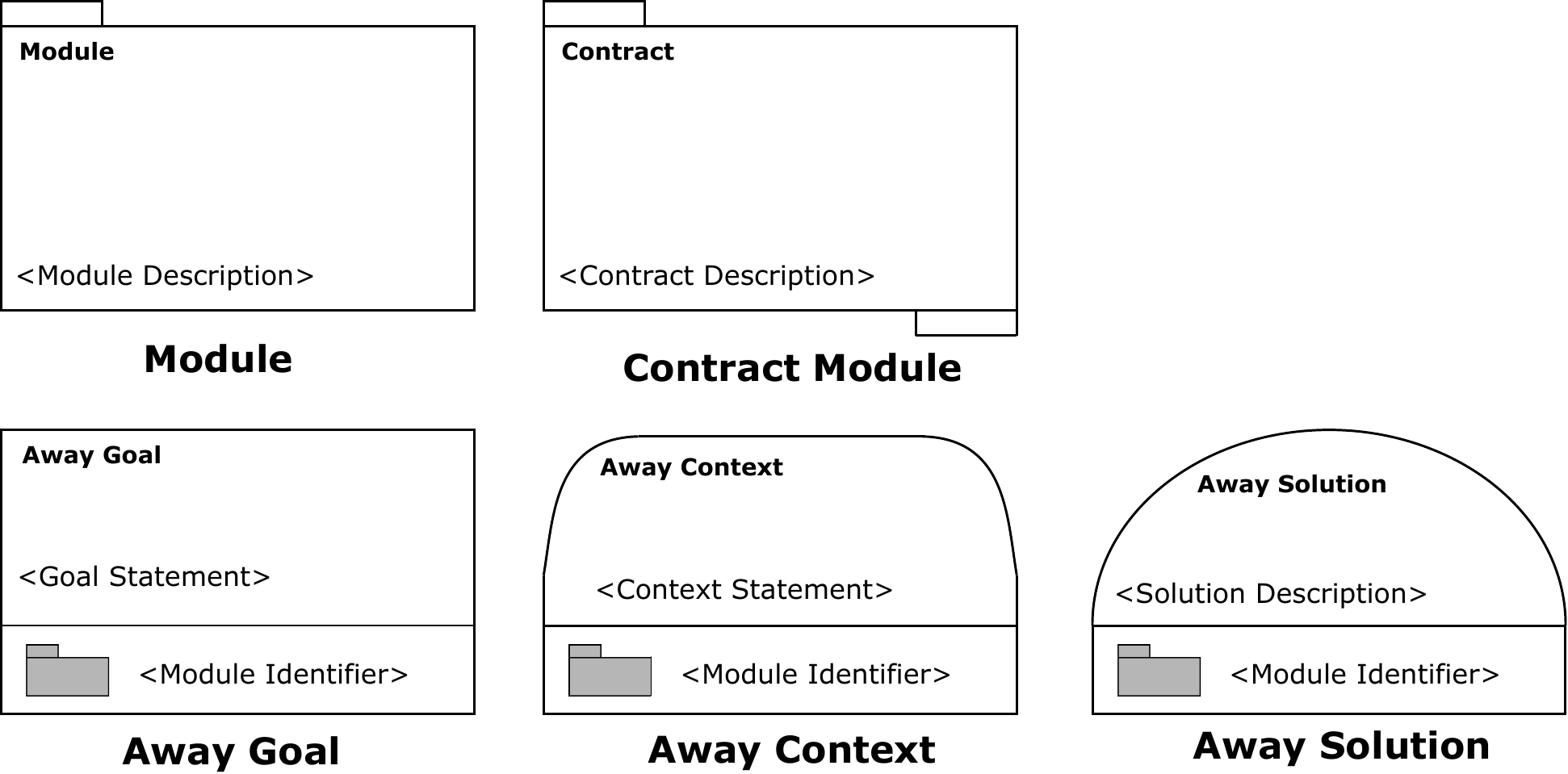}
	\caption{Modular GSN elements.}
	\label{fig:modularGSN}
\end{figure}

Goal structures can be organised in \textit{Module}s. 
For example, for a system that consists of two components A and B, it is possible to organise the safety cases of component A and B in two Modules \textit{MA} and \textit{MB}. 
Modularity promotes re-use, so that safety cases for system components can be re-used when different components are integrated to form a system. 
Figure~\ref{fig:modularGSN} shows the GSN elements that enable modularity support. 
When integrating system safety cases, a \textit{Contract Module} can be used to \textbf{bind} different \textit{Module}s together. 

\textbf{Binding} is done via the use of \textit{Away Goal}s, \textit{Away Context}s and \textit{Away Solution}, where \textit{Goal}s, \textit{Context}s and \textit{Solution}s from an external \textit{Module} can be referenced. 
Like other GSN elements, \textit{away} elements can be connected using \textit{SupportedBy} and \textit{InContextOf} connectors.

\begin{figure*}[htb]
    \centering
	\includegraphics[width=.7\linewidth]{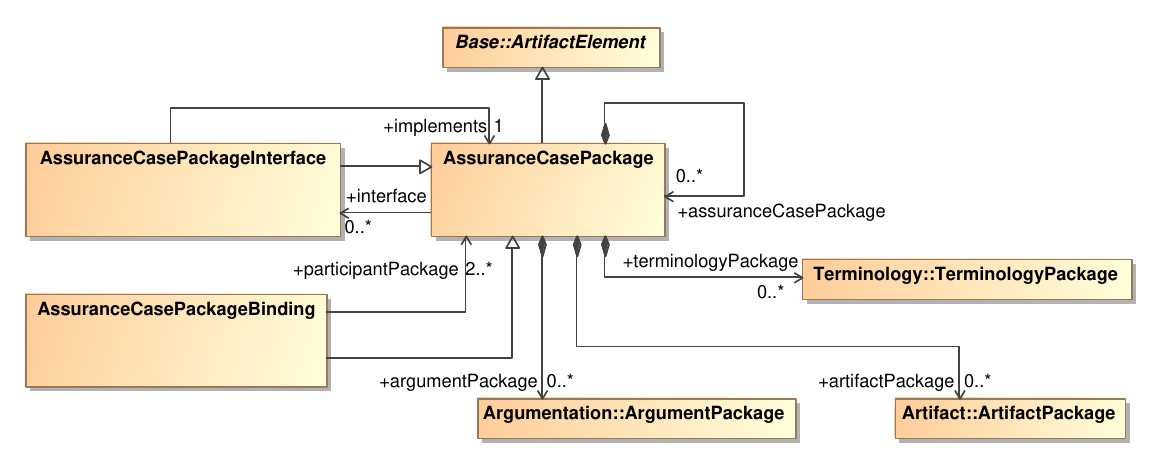}
	\caption{The Assurance Case component of SACM~\cite{sacm}.}
	\label{fig:sacm_ac}
\end{figure*}

\subsection{Model-Based Assurance Cases}
\label{sec_sacm}
Model Based Systems Engineering (MBSE)~\cite{brambilla2017model} is a contemporary systems engineering approach. 
In MBSE, \textit{model}s are first class artefacts, therefore \textit{driving} the development. 
MBSE has been proven to improve consistency and productivity significantly due to the automation provided by model management operations \cite{jaaksi2002developing, karna2009evaluating}. 

Over the past few years, model-based assurance case approaches emerged due to the benefits introduced by MBSE. 
Studies have shown how automated MBSE operations can be performed on model-based assurance cases (created using GSN) to check the well-formedness of assurance cases~\cite{denney2017tool}, generate and assemble structured argumentation within assurance cases~\cite{hawkins2015weaving}, and automatically generate texts for assurance case reports~\cite{denney2017tool}. 
However, existing model-based assurance case approaches (GSN and CAE - Cliams-Arguments-Evidence~\cite{bishop2000methodology}) do not provide sufficient support for traceability to engineering artifacts.
This is partly caused by the fact that GSN and CAE permit only structured arguments and not external artifact traceability.
This is a historical problem, as prior to model-based assurance case approaches, GSN and CAE are used to create physical documents, which naturally contain references to other (physical) engineering artifacts by their names.

\subsection{Structured Assurance Case Metamodel}
\label{sec:sacm}
Whilst graphical assurance case notations are powerful in expressing arguments regarding the safety of systems, they have their limitations. 
As discussed previously, an assurance case is not a self-contained document. 
That is, GSN elements (such as \textit{Context}s and \textit{Solution}s) may refer to engineering artifacts that provide contextual and evidential information.
Existing graphical notations (e.g. GSN and CAE) do not support such traceability. 

To address this limitation, the Object Management Group (OMG) specified and issued the \textit{Structured Assurance Case Metamodel} (SACM) \cite{sacm}. 
SACM is developed by the specifiers of existing system assurance approaches (e.g. GSN and CAE), based on the collective knowledge and experiences of safety and/or security practitioners over the period of last two decades. 
Therefore, features that are not previously supported by GSN and CAE have been evaluated and included in SACM. 

SACM organises model elements in \textit{Package}s to promote modularity, as shown in Figure~\ref{fig:sacm_ac}. 
An \textit{AssuranceCasePackage} may contain a number of \textit{TerminologyPackage}s (to store terms and expressions used in the assurance case), \textit{ArtifactPackage}s (to store artifacts, resources, events, etc. throughout the assurance case development process) and most importantly \textit{ArgumentPackage}s (to store safety/security arguments of a system or a component). 

\begin{figure*}[ht]
    \centering
	\includegraphics[width=.8\linewidth]{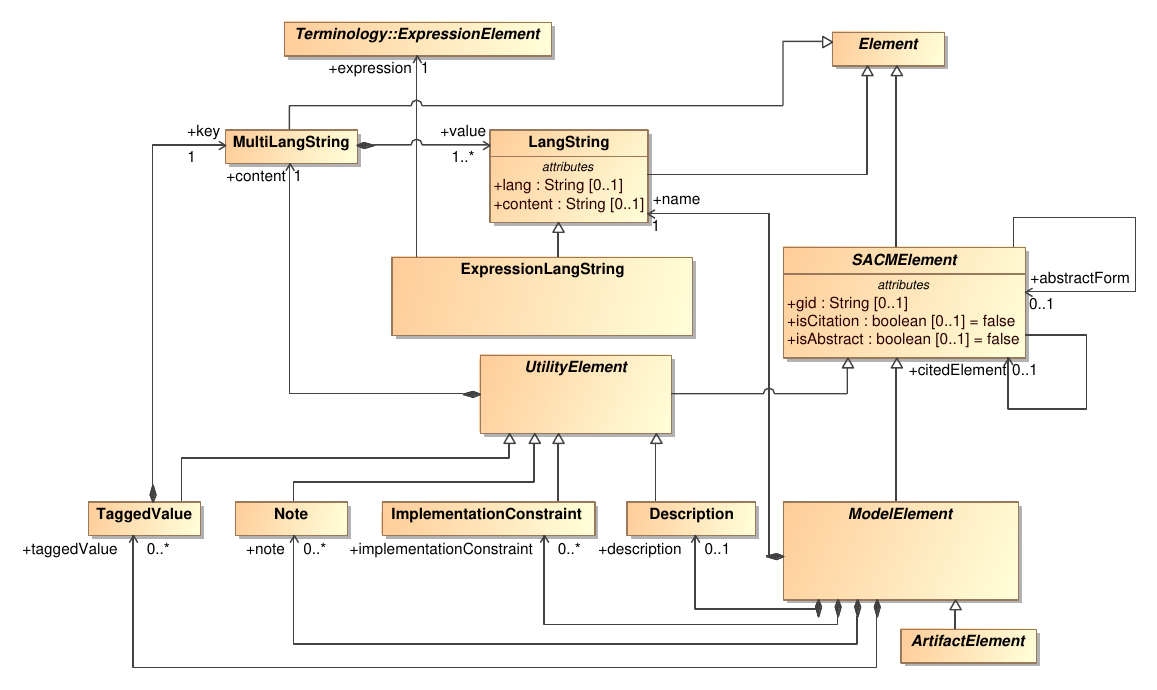}
	\caption{The Base component of SACM~\cite{sacm}.}
	\label{fig:sacm_base}
\end{figure*}

SACM provides essential concepts for complete model-based assurance cases (although currently there has not been approaches and tools to achieve it) in its \textit{Base} component shown in Figure~\ref{fig:sacm_base}.
For a \textit{ModelElement} in SACM, it can have a number of \textit{UtilityElement}s, in this work, we particularly focus on the \textit{ImplementationConstraint} concept, using which we describe the validation rules against engineering models.
In addition, it can also be seen that a \textit{ModelElement} can ``cite'' another \textit{SACMElement} via its \textit{CitedElement} association. 
This is a powerful mechanism, as it allows the users of SACM to cite any \textit{ModelElement} contained within one model.

\begin{figure*}[ht]
    \centering
	\includegraphics[width=.8\linewidth]{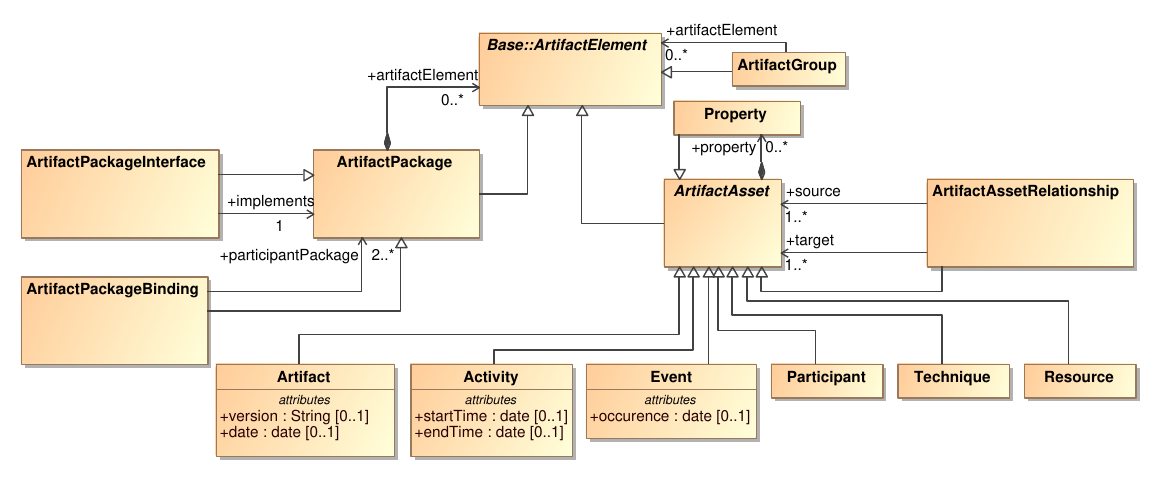}
	\caption{The Artifact component of SACM~\cite{sacm}.}
	\label{fig:sacm_artifact}
\end{figure*}

Another SACM component worth mentioning is the \textit{Artifact} component, as shown in Figure~\ref{fig:sacm_artifact}.
In this work, we make use of the \textit{ArtifactAsset}s (specifically, the \textit{Artifact} class) to demonstrate how we could record information (such as location, format and meta information) of external engineering artifacts and then use such information to perform automated verification and validation of such artifacts.

\subsection{Formal Methods and RoboChart}
\label{sec:formal}
Assurance cases can benefit from formal methods~\cite{Gleirscher2018-NewOpportunitiesIntegrated}. 
Informal safety arguments and evidence can be difficult to automatically evaluate, and may be subject to some argumentation fallacies~\cite{greenwell2006taxonomy}.
Thus, formalisation of requirements, to allow the use of formal methods, can significantly improve both automation and the confidence~\cite{foster2020formal}. 
At the same time, assurance cases allow us to put formal methods results in context. 
For systems assurance it is never enough to simply prove properties of a formal model. 
For the results to be meaningful, the model must be linked to its corresponding real-world artifact~\cite{Lee2018}, such as software and hardware by some form of validation argument.
Consequently, a comprehensive demonstration of safety requires both assurance cases and formal methods.

An important development here is Isabelle~\cite{Isabelle}, a verification framework for integrated formal methods~\cite{Wenzel2007FMIsabelle,Wenzel2019-Isar,Foster2019-iFM}. 
Development centres around documents called Isabelle \textit{theories}, which encode graphs of hyperlinked mathematical artifacts, such as definitions, theorems, and proofs. 
Formal method integration is supported by (1) a flexible front-end, which supports a variety of languages~\cite{Tuong2019-CIsabelle} and their translation into formal semantics; (2) an extensible plugin-oriented architecture where external tools, such as SMT solvers~\cite{Blanchette2011}, can improve automation; and (3) incremental theory processing~\cite{Wenzel2007FMIsabelle}. 
Moreover, Isabelle can be installed as a server component which other tools can make use of as a verification tool service.
At the foundational level, Isabelle can be used for mechanisation of a variety of formal semantics~\cite{Nipkow2014-ConcreteSemantics}, to support verification tools. 
A front-end for a programming language, such as C~\cite{Tuong2019-CIsabelle}, can be developed to support concrete program verification~\cite{Alkassar2008}. 
Moreover, SACM was recently implemented in Isabelle to create Isabelle/SACM~\cite{Foster2019-iFM}, which allows verification of assurance cases, integration with formal evidence, and generation of certification documents~\cite{Brucker2019-DOFCert}.

Formal methods can be difficult for non-experts to apply, and so there is a desire to use model-based graphical frontends.
For example, the RoboChart language~\cite{Miyazawa2019-RoboChart} is a graphical language for the architectural and behavioural description of a robotic controller. 
It includes a formalised subset of the UML state machine notation with a complete formal semantics in the CSP process algebra~\cite{Brookes1984}. The formal semantics allows RoboChart models to be subjected to formal analysis using model checking~\cite{miyazawa2016robochart,Miyazawa2019-RoboChart} and theorem proving~\cite{Foster2018-FACS,foster2020formal}. RoboChart is therefore both accessible to practitioners, and at the same time uses formal methods to allow development at higher assurance levels.

A RoboChart model consists several elements:

\begin{enumerate}
    \item a \emph{data model}, consisting of data types and functions with pre- and postconditions;
    \item \emph{interfaces}, which collect together variables, events, operations, and clocks that can be used by other components;
    \item \emph{robotic platforms}, which abstract the hardware by providing variables and events, potentially through provided interfaces;
    \item \emph{controllers}, which describe different units that control the robot and communicate with other controllers and the robotic platform using required interfaces;
    \item \emph{state machines}, which are used to describe the behaviour of controllers and can communicate using events and shared variables;
    \item an \emph{architectural model}, which describes the connections between controllers and robotic platforms.
\end{enumerate}

States and transitions in a RoboChart state machine can specify actions using a formal action language inspired by CSP. It includes primitives for receiving events $c?v$, sending them $d!e$, and assigning values to variables $x := e$. RoboChart also has a discrete time model, and each controller can share a number of clocks that can be used to observe the passage of time. 

In this paper, we use RoboChart for modelling an Autonomous Underwater Vehicle (AUV).
To support such languages, particularly where there is a diversity of such notations, an integrated approach is required. 
Model-based assurance cases allow such an integration of heterogeneous models, and support justification and traceability for these models, and any associated analysis results. 

%% file: section_3.tex
\begin{figure*}[ht!]
    \centering
    \includegraphics[width=1\linewidth]{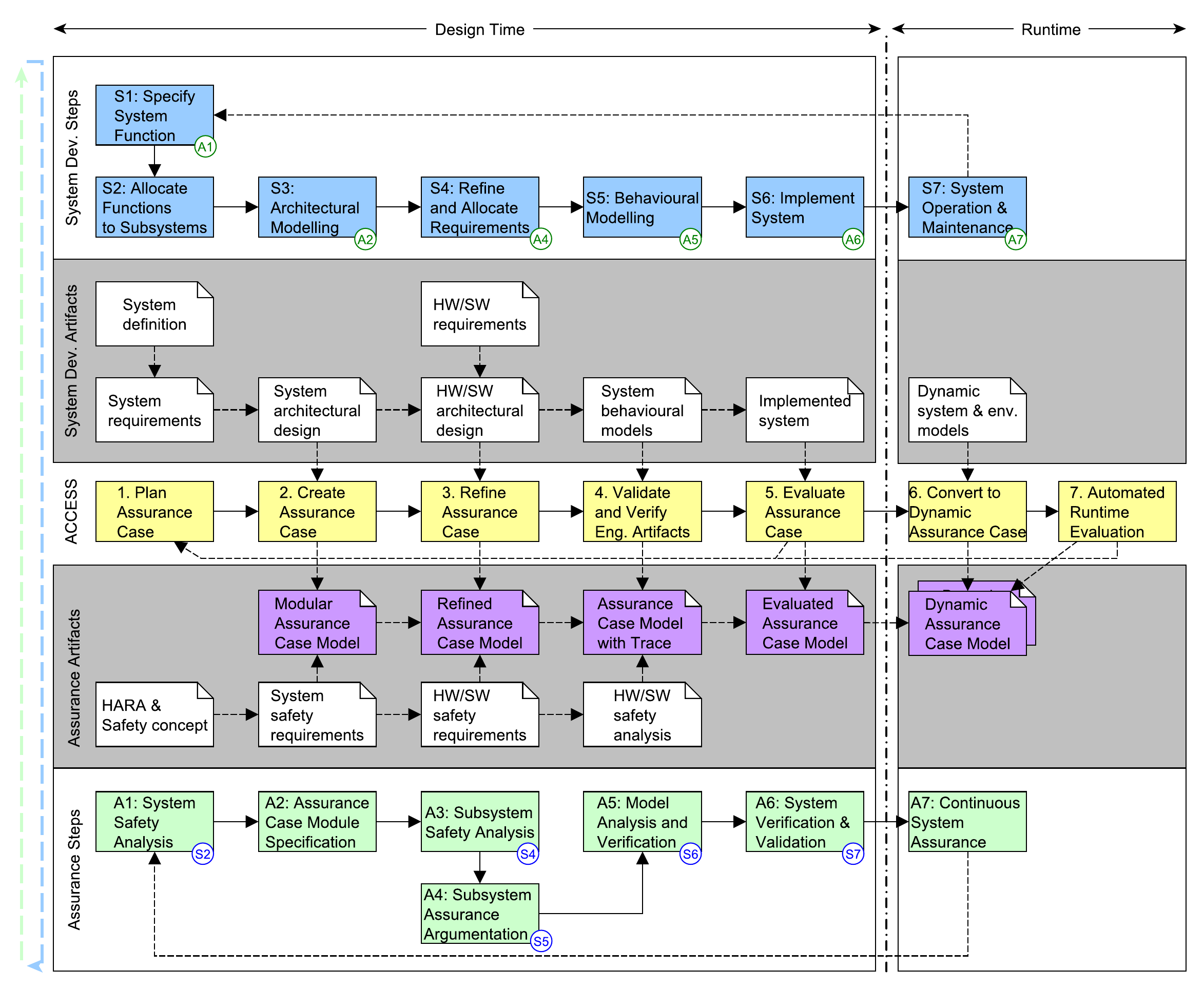}
    \caption{ACCESS Process Overview}
    \label{fig:access}
\end{figure*}

\section{Approach Overview}
\label{sec:access}

In this section, we propose the methodology of \textit{Assurance Case Centric Engineering of Safety-critical Systems} --ACCESS, which is an engineering methodology for developing and assuring a system (including both its hardware and software) around an evolving Assurance Case, adopting principles of \textit{Model Based Systems Engineering} (MBSE).
We use the term ``model'' in a broad sense to encompass any structured machine-readable artifacts, which include resources like EMF-based models, XML files, spreadsheets, databases as well as models created using other technologies (e.g. Simulink).

The ACCESS process is illustrated in Figure~\ref{fig:access}. 
We consider activities in both the System Development Process (boxes rendered in blue in the upper swim lane with white background) and the System Assurance Process (boxes rendered in green in the lower swim lane with white background).
For each kind of the processes, we identify a key set of engineering artifacts (swim lanes with gray background), and we discuss the relationships between the engineering artifacts and the assurance case of the system.

There are 7 steps in the ACCESS methodology (rendered in yellow), each step coordinates System Development activities and System Assurance activities. 
The columns in Figure~\ref{fig:access} indicates the scopes of ACCESS steps.
Within each ACCESS step, activities can be iterative (this is indicated by the circular dashed arrow lines on the left side of the swim lanes), that is, System Development and System Assurance activities can be repeated until each ACCESS step is deemed sufficiently executed.
Activities in both groups are interleaved, in Figure~\ref{fig:access}, there are circles at the bottom right corner of some of the activities, indicating that practitioners are advised to continue the development by conducting the activity identified in the circles (e.g. S1 to A1).

\subsection{Step 1: Plan Assurance Case}
In this step, the first task to perform is \textbf{S1: Specify System Function}, in which the high-level system functional requirements are defined, the hardware platform is chosen, and assumptions of the environment are specified. 
The function specification provides the top-level contract for the system: provided it is deployed in an environment satisfying the assumptions, it will perform the required functions. 
Once this task is complete, the assurance process begins in task (\textbf{A1: System Safety Analysis}), which includes activities such as Hazard Analysis and Risk Assessment (HARA), using analysis approaches such as Failure Mode and Effects Analysis, Fault Tree Analysis, etc..
From the analysis, a preliminary list of \textit{Safety Goals} can be derived, forming the \textit{Safety Concept} of the system.
Based on the safety concept, system development task \textbf{S2: Allocate Functions to Subsystems} shall be performed, defining subsystems, and the interface between them. 

Outcome of this step may include: system definition, system requirements, HARA and the Safety concept.

\subsection{Step 2: Create Assurance Case}
In task \textbf{S3: Architectural Modelling} the architecture of the system is modelled, including subsystem blocks, functionalities provided by the hardware platform, and connections between the various components. 
Based on this architecture model, in task \textbf{A2: Assurance Case Module Specification}, corresponding assurance case modules shall be specified. 
This includes generation of a public claim for each requirement that has been allocated to a particular subsystem, each of which needs an associated argument, and also public assumptions that will need to be satisfied by peer subsystems. 

Outcome of this step may include: system architectural design, system safety requirements and a modular assurance case model.

\subsection{Step 3: Refine Assurance Case}
The draft assurance case created in ACCESS step 2 is further refined. 
In this step, the first task to perform is \textbf{A3: Subsystem Safety Analysis}, in which safety analysis is performed for every subsystem identified in the system architectural design. 
Next, in task \textbf{S4: Refine and Allocate Requirements}, system requirements (as well as safety requirements) are allocated to subsystems, this requirement allocation shall be preferably traceable.
Then, task \textbf{A4: Subsystem Assurance Argumentation} is performed, in which safety arguments are developed for each of the subsystem, it is to be noted that the argumentation shall correspond to the system requirements and safety requirements, and traceability shall be maintained.
Tasks in this step shall be performed iteratively until the refined assurance case model is deemed sufficiently mature.

Outcome of this step may include: hardware/software requirements, architectural design, safety requirements and the refined assurance case model.

\subsection{Step 4: Validate and Verify Engineering Artifacts} 
With the assurance case model in place, the next step is to investigate how it could be verified and validated, and preferably in an automated manner.
In our process, we introduce an additional task, \textbf{S5: Behavioural Modelling}, which is typical for software development process.
This task can include the creation of state machines or sequence diagrams for each of the subsystem, conforming to the architectural model.
We assume that the behavioural modelling notation will have a formal semantics, as is the case for example with the RoboChart language~\cite{Miyazawa2019-RoboChart}. 
Next, we propose the task of \textbf{A5: Model Analysis and Verification}. 
The purpose here is to ensure that each of the requirements is represented in or satisfied by the behavioural model, and reflected in the assurance case. 
For example, we may wish to state that a particular technical requirement is implemented by a transition, or that a state machine satisfies a global safety requirement. 
We may also wish to verify the integrity of the model, for example by checking for the possibility of deadlock or livelock using a model checker. 
This task may expose flaws in the behavioural model, such as inadequately implemented requirements, and so there is iteration between \textbf{S5} and \textbf{A5}, which should in all circumstances be carried out by separate teams.

In this step it is possible to establish traceable links from the assurance case to its supporting models to form a self-contained assurance case model repository, this is where benefits of MBSE emerge -- that some (if not all) of the verification and validation tasks can be automated from the assurance case as an entry point.

Outcome of this step may include: System behavioural model, hardware/software safety analysis and an assurance case model with trace.

\subsection{Step 5: Evaluate Assurance Case}
Once a robust behavioural model is created, where every requirement can be traced to a model element, or verification property, the design can be synthesised into an implementation by task \textbf{S6: Implement System}. 
This task can invoke a variety of techniques, including formal code and data refinement, code generation, and manual coding. 
In parallel with implementation, the assurance process executes task \textbf{A6: System Verification and Validation}, all hardware and software components are verified based on the requirements (for both system and safety) using techniques such as model-based testing and code verification, validation activities are also performed to support the validity of the assurance case.
Again, these two tasks are iterative as verification may expose implementation or design issues that need addressing. 

Unsuccessful verification results are reflected in the assurance case (presumably in an automated fashion), and shall be fixed by repeating previous ACCESS steps where necessary. 
The assurance case is ``complete'' once no unsupported claims remain. 
The assurance case will also need to be independently evaluated, though this is not shown in the process as evaluation is cross-cutting in all activities.
Presumably, assurance case evaluation is automated provided that an MBSE approach is adopted and the traceablility links from the assurance case to its supporting engineering artifacts are established.

Outcome of this step may include: implemented system, system verification reports, system validation reports.

\subsection{Step 6: Convert to Dynamic Assurance Case}
For RASs, it is becoming imperative to argue the safety at runtime. 
Therefore, this (optional) step targets system operation and maintenance.
In task \textbf{S7: System Operation and Maintenance}, requirements for the safe operation and maintenance of the system are specified, including safety related requirements (e.g. safe operation protocols, safe maintenance procedures). 
In task \textbf{A7: Continuous System Assurance}, practitioners shall determine, based on system operation and maintenance context, which part of the assurance case shall be converted to dynamic (i.e. that runtime data are reflected to engineering models at runtime) one, so that assurance can still be carried out at system runtime.

Outcome of this step includes (but not limited to): Dynamic system models, system environment models and dynamic assurance case model.

\subsection{Step 7: Automated Runtime Evaluation}
Once the a dynamic assurance case is obtained, in this step is to perform automated runtime evaluation, to automatically evaluate the parts of the assurance (and its supporting engineering artifacts) to constantly check the validity of the assurance case.
It is to be noted that the assurance case evaluation shall be non-invasive, that it only provides the system with the validity of the assurance case, and shall not take control of the system in any form.

The evaluation results can be used by the system at runtime to yield system safety status at runtime, and take measures to get back to the safe state.
Evaluation results shall also be recorded and reviewed for continuous improvement of the system, and ACCESS steps can be repeated for this purpose.

%% file: section_4.tex
\section{Tool Support}
\label{sec:toolsupport}

\begin{figure}[h]
    \centering
	\includegraphics[width=.7\linewidth]{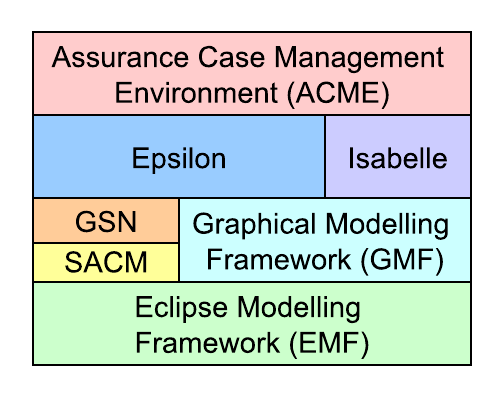}
	\caption{Assurance Case Management Environment (ACME).} 
	\label{fig:acme_arch}
\end{figure}

The ACCESS methodology is backed by our tool support -  \textit{Assurance Case Management Environment} (ACME).
ACME is an integrated model based assurance case modelling and management framework, that supports the creation and management of assurance case models that conform to the Structured Assurance Case Metamodel (SACM)~\cite{sacm}.
Since SACM is relatively new and its graphical notations are being standardised, ACME also supports the Goal Structuring Notation (GSN), whose current model-based implementation extends the abstract syntax of SACM, explained in detail in~\cite{wei2019model}.
In this way, ACME supports the creation of model-based GSN diagrams, and at the same time provides access to all other features of SACM.

The architecture of ACME is illustrated in Figure~\ref{fig:acme_arch}.
We implement SACM and GSN with the Eclipse Modelling Framework (EMF)~\cite{steinberg2008emf}, and use Graphical Modelling Framework (GMF)~\cite{gmf} to create graphical editors for SACM packages and GSN modules.s
To enable automated model management and the checking of formal notations, we also integrate:
\begin{itemize}
    \item The Eclipse Epsilon platform~\cite{kolovos2008epsilon}, which is an integrated model management platform, that provides task specific model management languages (model validation, model transformations, etc.) that operate on models defined in different modelling technologies (EMF, Excel spreadsheet, Simulink~\cite{simulink}, UML, etc.);
    \item Isabelle~\cite{Isabelle}, which is a generic proof assistant that allows mathematical formulas to be expressed in a formal language and provides tools for proving these formulas in a logical calculus, with a high degree of automation. 
    Isabelle also provides an extensible document model which can support the encoding of different meta-models and associated parsers.
\end{itemize}
Using SACM's full potential and with the help of model management frameworks, ACME currently supports 1) fine-grained traceability from an assurance case to its referenced engineering artifacts (defined in mainstream modelling technologies) to the level of model element(s); 2) traceability to formal notations in Isabelle; and 3) automated means to validate/verify traced engineering artifacts. 
It is to be noted that ACME also support other high-level functionalities, but they are not within the scope of this paper.

To enable the traceability from an assurance case to its supporting engineering artifacts, we make use of SACM's \textit{Artifact} component (discussed in Section~\ref{sec_sacm}) in ACME.
With the help of the \textit{Artifact} component of SACM, in an \textbf{Artifact} element, we are able to record inside it: \textbf{(1)} the type of an engineering artifact (e.g. EMF model, XML document, Excel spreadsheet, Isabelle theory file, etc.) to trace to; \textbf{(2)} the location of the engineering artifact; and \textbf{(3)} the meta-data of the engineering artifact (e.g. metamodel, XML metadata, etc.).
To obtain a more fine-grained traceability, we make use of the \textbf{ImplementationConstraint} (discussed in Section~\ref{sec_sacm}) element, and record model querying/validation programs written in a model querying language (in this work we support the Epsilon Object Language (EOL)~\cite{kolovos2006epsilon}, but any other languages can be supported) in \textbf{Artifact} model elements.
In this way, when the program is executed, we are able to obtain specific model element(s) (or information) from the engineering artifact that an \textit{Artifact} element refers to, which can be used in assurance cases to provide contextual and evidential information.
In summary, traceability from an assurance case to external engineering artifacts is achieved by referring to \textbf{Artifact} elements (organised in an \textbf{ArtifactPackage}) that contain traces to engineering models, from (in GSN terms) either an \textbf{Context} or an \textbf{Solution} element organised in a GSN \textbf{Module}. 

With the traceability from an assurance case to engineering artifacts, we are also able to perform automated assurance case evaluations (i.e. validation and verification on referenced external engineering artifacts).
For validation, we refer to external engineering artifacts using model elements defined in SACM's \textit{Artifact} component, and embed programs such as validation rules (which return \textit{true} or \textit{false}). 
ACME executes the validation rules and reflects the validation results to ACME editors so that the users can find out which part of an assurance case failed in the evaluation.
For formal verification, we refer to Isabelle \textit{theory documents}s. 
A theory document is a hierarchical structure consisting of formal artifacts, such as data types, functions, theorems, and proofs. 
Upon execution, ACME sends the \textit{theory file}s to the established Isabelle server (discussed in~\cite{Wenzel2019-Isar}), which can be communicated with using an RESTful API.
When a theory file is sent to the Isabelle server, the server processes the file and returns JSON messages conveying the status of all artifacts contained within the theory file.
If the processing of any artifacts fail, the JSON messages contain all the problems that Isabelle found. 
For example, a candidate proof could fail to prove a theorem, and this would raise an error.
In ACME, we trace to an Isabelle theory file with an \textbf{Artifact} and perform formal notation checking in an automated manner.
If an Artifact cannot be verified, ACME reflects this information in the model editor. 
Moreover, ACME also supports the translation of an assurance case to an Isabelle/SACM theory~\cite{Foster2019-iFM}, which can be used for verifying its logical integrity. 
If there are any errors in other assurance case nodes, these are likewise reflected.

\begin{figure}[h]
    \centering
	\includegraphics[width=1\linewidth]{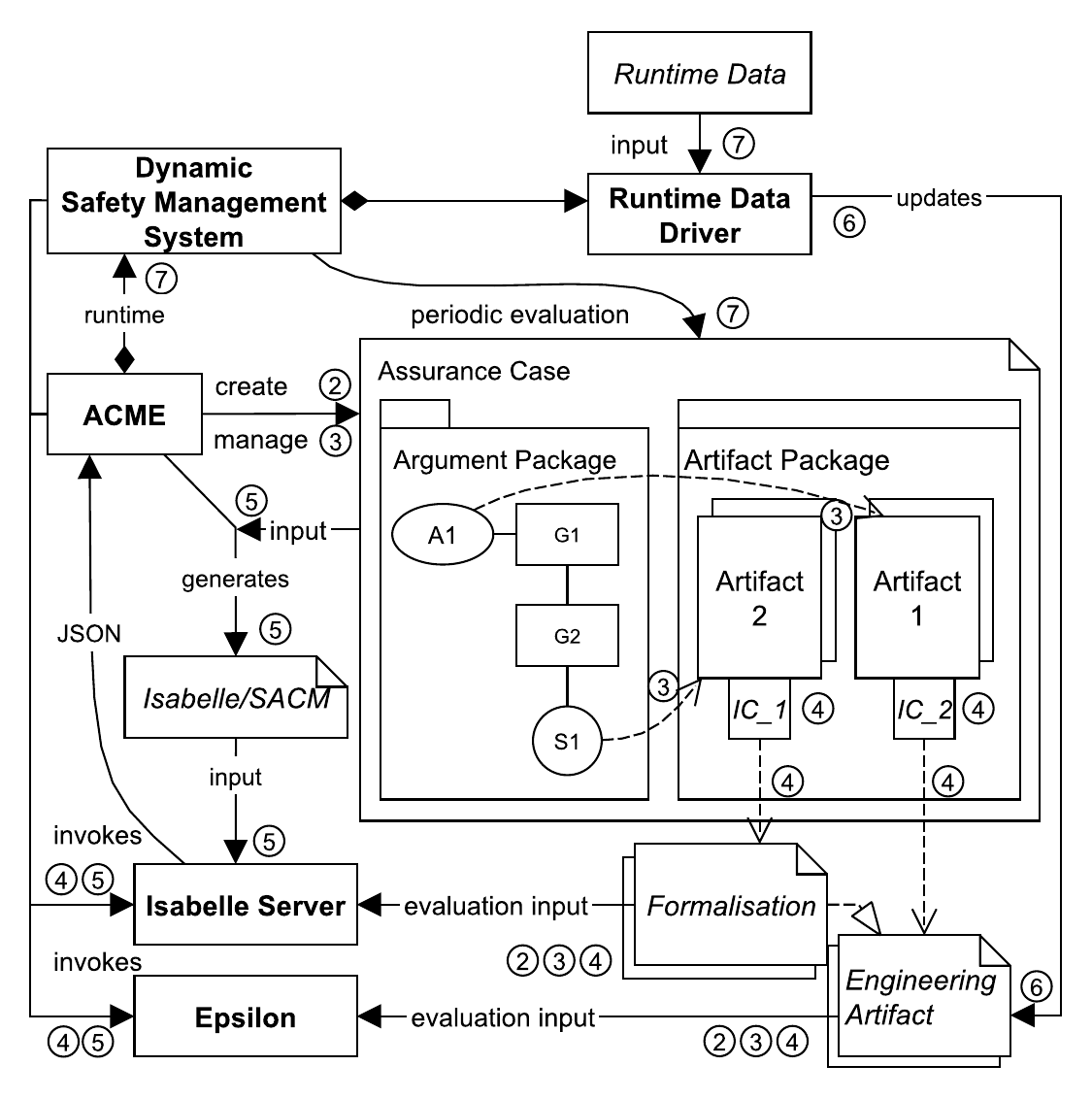}
	\caption{ACME workflow in the context of ACCESS.}
	\label{fig:approach}
\end{figure}

The support for the ACCESS methodology from ACME is illustrated in Figure~\ref{fig:approach}. 
Whilst in this section we illustrate the support from ACME, we argue that any model-based assurance case management framework may apply the ACCESS methodology to develop safety critical systems.

For ACCESS \textbf{Step 2} and \textbf{Step 3}, we use ACME to create and manage a model-based assurance case, which may contain a number of \textbf{ArgumentPackage}s, \textbf{TerminologyPackage}s and \textbf{ArtifactPackage}s. 
In the figure we show how elements inside \textbf{ArgumentPackage}s can link to elements in \textbf{ArtifactPackage}s. 

In ACCESS \textbf{Step} \textbf{2}, \textbf{3} and \textbf{4}, various engineering artifacts (such as requirement models, architecture models, safety analysis models, and behavioural models) are produced, and reside alongside the assurance case (these are shown on the bottom right corner).
\footnote{In this paper we focus only on Engineering Artifacts that can be automatically validated with Epsilon, and Formalisations that can be automatically verified by Isabelle.}

In ACCESS \textbf{Step} \textbf{4}, the evaluation of engineering artifacts can be automated within ACME, with the assurance case as the entry point.
With SACM's \textbf{ImplementationConstraint} (IC) model element, we create validation rules written in EOL (\textit{IC\_1 and IC\_2 in Figure~\ref{fig:approach}}) for the referenced artifacts. 
Based on the type of the engineering artifacts, ACME determines if it should invoke Epsilon or Isabelle. 
The results of the evaluations are processed by ACME, and if any problem occurs, they will be marked on the \textit{Artifact}s that contain the evaluation rules in the ACME editor.

In ACCESS \textbf{Step 5}, we perform the evaluation of the entire assurance case, which includes the verification and validation on the system level. 
Using ACME, we perform two types of evaluation.
The first type is to invoke evaluation on all referenced engineering artifacts from the assurance case. 
In ACME we provide an \textit{evaluate} function which can be called on an assurance case, ACME then automatically looks into all \textbf{ArgumentPackage}s and looks for argument elements that refer to artifact elements in \textbf{ArtifactPackage}s, then ACME automatically calls Isabelle or Epsilon and determines if all evidence (that support the assurance case) are valid.
The second type is to formalise the assurance case and check its logical integrity in an automated manner.
We do this by transforming the assurance case to Isabelle/SACM notations with ACME's built-in model-to-text transformation (written in the Epsilon Generation Language~\cite{rose2008epsilon} -- EGL) and generate the Isabelle/SACM notation representation of the assurance case in the form of a \textit{theory file}.
The Isabelle/SACM notation is then sent to the Isabelle server for machine-checking. 
The evaluation result will be parsed by ACME, which then locates the model elements in the assurance case that fail the evaluation, and displays their corresponding error messages.

In ACCESS \textbf{Step 6}, elements of the assurance case can be converted to dynamic ones, whose validity depend on runtime data.
In order to do this, ACME allows the creation of \textit{Runtime Data Driver}s, which provide connections between runtime data and engineering artifacts.
In this way, runtime data can be constantly reflected to engineering artifacts.

In ACCESS \textbf{Step 7}, ACME's runtime component -- \textit{Dynamic Safety Management System} (\textit{DSMS}) is used to evaluate the assurance case. 
\textit{DSMS} performs automated periodic evaluation on the assurance case, it does so by automatically invoking validation rules on dynamic parts of the assurance case (which refer to \textit{Engineering Artifact}s that are updated by \textit{Runtime Data Drivers} at runtime).
In this way, we can assure the assumptions about the system (e.g. its behaviour or its operational environment), check dynamic evidence, and determine operation contexts, which form the baseline for dynamic assurance case evaluation.

%% file: section_5.tex
\section{Case Study}
\label{sec:case}

\begin{figure}[!ht]
    \centering
	\includegraphics[width=.8\linewidth]{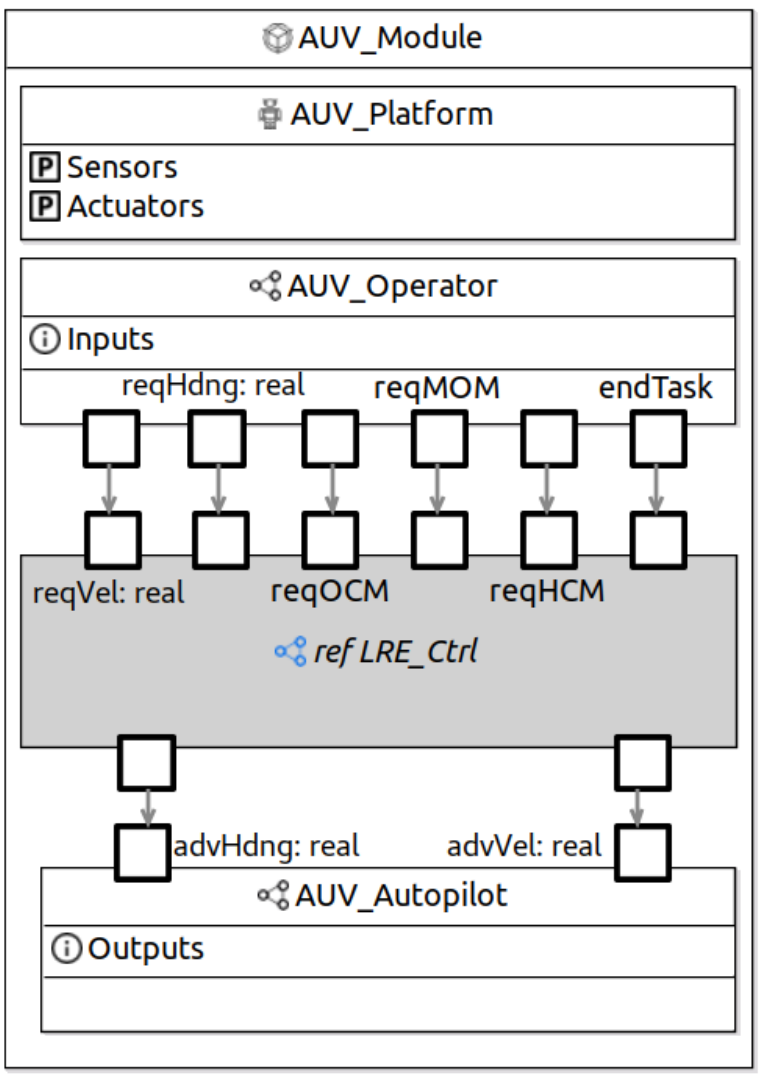}
	\caption{Overall Architecture of the AUV.}
	\label{fig:auv_architecture}
\end{figure}

\begin{figure}[!h]
    \centering
	\includegraphics[width=1\linewidth]{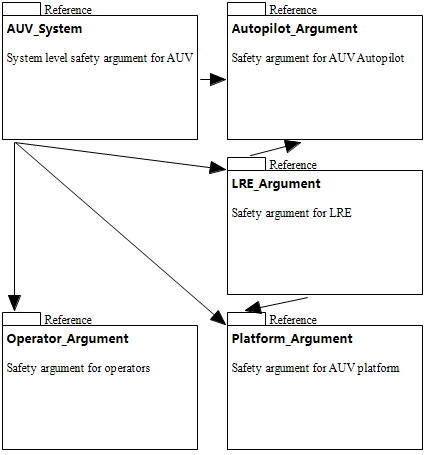}
	\caption{Overall AUV safety argument structure.}
	\label{fig:auv_overall}
\end{figure}

In this section, we evaluate the ACCESS methodology by applying it to a development process of an Autonomous Underwater Vehicle (AUV).
The assurance case for the AUV is developed based on an integrated RoboChart~\cite{miyazawa2016robochart} model, which includes the architecture of the AUV and the behaviour of its controllers, discussed in~\cite{foster2020formal}.
We also show that it is possible, with the help of ACME, to automate the evaluation of model based assurance cases and have the assurance case \textit{drive} the development of the system.

It is to be noted that due to confidentiality and the complexity of the AUV, in this case study we only demonstrate activities in the system development and system assurance of ACCESS where the benefits of automation can be reflected.

\subsection{ACCESS Step 1}
The AUV is a portable untethered remotely operated vehicle, equipped with a visual mapping system and verified on-board autonomy.
The aim is to make it capable of conducting light intervention tasks, such as cathodic protection surveys (oil and gas) and simple coring (offshore), with potential to move to more complex interventions in a later phase, such as valve turning.
The project brings together the UK expertise from: the National Oceanography Centre and Forth Engineering in Underwater Robotic Development; ROVCO on subsea operation, sensor development and subsea vision preception; and D-RisQ in Software Verification.

The National Oceanography Centre engages with regulators through their ongoing contribution to the Marine Autonomous Systems regulatory working group to ensure regulatory compliance. 
To this end, the use of a structured assurance case is vital to communicate the evidence of safety operation to non-specialists, especially in the aspect of software controlled autonomous behaviour.

In \textbf{ACCESS Step 1}, system functions are specified (task \textbf{S1}) and safety analysis on the system level is performed (task \textbf{A1}), and functions of the subsystems are allocated (task \textbf{S2}).
We do not show the outputs of these tasks in detail due to confidentiality, but it is to be noted that the requirements are also model-based.

\subsection{ACCESS Step 2}
The overall architecture of the AUV is modelled (task \textbf{S3}) using the RoboChart language~\cite{miyazawa2016robochart}, shown in Figure~\ref{fig:auv_architecture}. 
The robotic platform (\textit{AUV\_Platform}) acts as an abstraction layer for the hardware, and provides shared variables for sensors, actuators and events. 
The operator, which can be a human or navigation system, provides instructions to the LRE (\textit{LRE\_Ctrl} - Last Response Engine) to support execution of tasks, such as requesting a particular heading and velocity.
The LRE sits between the operator (\textit{AUV\_Operator}) and the autopilot component (\textit{AUV\_Autopilot}).
The LRE's job is to avoid hazardous behaviours, such as getting too close to an obstacle, or entering \textit{Object Proximity Exclusion Zone}s (OPEZs), and engaging evasive manoeuvres if necessary.
The autopilot controls the AUV actuators, and takes advice only from the LRE.

Based on the architectural design, we create a modular assurance case (task \textbf{A2}) for the AUV, as shwon in Figure~\ref{fig:auv_overall}.
It contains 5 \textit{argument package}s (see Section~\ref{sec_sacm}), which are represented as GSN modules.
The \textit{AUV\_system} module contains the argument of system level safety for the AUV, including hazard analysis and allocation of safety requirements. 
It is supported by modules \textit{Platform\_Argument}, \textit{Operator\_Argument}, \textit{LRE\_Argument} and \textit{Autopilot\_Argument}.
This means that the validity of \textit{AUV\_System} depends on the validity of all 4 modules that support it.
In addition, \textit{LRE\_Argument} depends on \textit{Platform\_Argument} and \textit{Autopilot\_Argument}.

\begin{table*}[h]
\centering
\resizebox{\textwidth}{!}{%
\begin{tabular}{|c|c|c|c|c|c|c|c|c|}
\hline
\textbf{\begin{tabular}[c]{@{}c@{}}Component\\ ID\end{tabular}} & \textbf{\begin{tabular}[c]{@{}c@{}}Failure\\ Rate\end{tabular}} & \textbf{\begin{tabular}[c]{@{}c@{}}Safety\\ Related\end{tabular}} & \textbf{\begin{tabular}[c]{@{}c@{}}Failure\\ Mode\end{tabular}} & \textbf{\begin{tabular}[c]{@{}c@{}}Failure\\ Mode\\ Distribution\end{tabular}} & \textbf{\begin{tabular}[c]{@{}c@{}}Safety\\ Goal\\ Violation\end{tabular}} & \textbf{\begin{tabular}[c]{@{}c@{}}Safety \\ Mechanism\end{tabular}} & \textbf{\begin{tabular}[c]{@{}c@{}}Failure Mode\\ Coverage by\\ Safety Mechanism\end{tabular}} & \textbf{SPF/RF} \\ \hline
D1                                                              & 10                                                              & Yes                                                               & Open                                                            & 30\%                                                                           & Yes                                                                        & None                                                                 & 0\%                                                                                            & 3            \\ \hline
                                                                &                                                                 &                                                                   & Short                                                           & 70\%                                                                           &                                                                            &                                                                      &                                                                                                &              \\ \hline
C1                                                              & 2                                                               & Yes                                                               & Open                                                            & 30\%                                                                           &                                                                            &                                                                      &                                                                                                &              \\ \hline
                                                                &                                                                 &                                                                   & Short                                                           & 70\%                                                                           &                                                                            &                                                                      &                                                                                                &              \\ \hline
C2                                                              & 2                                                               & Yes                                                               & Open                                                            & 30\%                                                                           &                                                                            &                                                                      &                                                                                                &              \\ \hline
                                                                &                                                                 &                                                                   & Short                                                           & 70\%                                                                           &                                                                            &                                                                      &                                                                                                &              \\ \hline
L1                                                              & 15                                                              & Yes                                                               & Open                                                            & 30\%                                                                           & Yes                                                                        & None                                                                 & 0\%                                                                                            & 4.5          \\ \hline
                                                                &                                                                 &                                                                   & Short                                                           & 70\%                                                                           &                                                                            &                                                                      &                                                                                                &              \\ \hline
R1                                                              & 1                                                               & No                                                                & Open                                                            & 30\%                                                                           &                                                                            &                                                                      &                                                                                                &              \\ \hline
                                                                &                                                                 &                                                                   & Short                                                           & 70\%                                                                           &                                                                            &                                                                      &                                                                                                &              \\ \hline
Lamp1                                                           & 150                                                             & No                                                                & Open                                                            & 100\%                                                                          &                                                                            &                                                                      &                                                                                                &              \\ \hline
U1                                                              & 100                                                             & Yes                                                               & RAM                                                             & 100\%                                                                          & Yes                                                                        & ECC                                                                  & 99\%                                                                                           & 1            \\ \hline
\end{tabular}%
}
\caption{Fragment of the Failure Mode and Effect Diagnostic Analysis (FMEDA) for the AUV.}
\label{tab:fmeda}
\end{table*}

\subsection{ACCESS Step 3}
We then perform safety analysis on subsystems (task \textbf{A3}).
In critical systems development, it is often required that inductive and deductive safety analysis to be performed.
One typical analysis that is performed frequently in the development of safety critical systems is the Failure Mode and Effect Diagnostic Analysis (FMEDA). 
FMEDA looks at the failure mode, failure mode distribution and failure rate of system components (from simple components such as capacitors to more complex components such as Microcontrol Units), as well as the safety mechanisms to prevent failures, in order to compute hardware design metrics (e.g. Single Point Failure Metrics - SPFM) to determine the safety integrity levels of components.
In this case study we demonstrate how we could establish traces to FMEDA from the assurance case in ACME, and automatically validate the assurance case using the FMEDA results as evidence.

ACME allows the traceability to engineering artifacts defined in arbitrary modelling technologies.
One particular type of engineering artifact is Excel spreadsheets, which is often used in FMEDA. 
Table~\ref{tab:fmeda} shows a fragment of the FMEDA performed on the power supply of the proximity sensor (in which D1 is a Diode, C1 and C2 are capacitors, L1 is an inductance, R1 is a resistor, Lamp1 a lamp and U1 a Microcontrol Unit). 
Note that the unit of \textit{Failure Rate} is \textit{Failures in Time} (FIT), which is $10^{-9}$ times/hour, where SPF stands for Single Point Faults and RF stands for Residual Faults (faults that are not covered by safety mechanisms).
With the FMEDA, it is possible to compute the SPFM of a particular hardware component, using the formula:

$$SPFM = 1 - \frac{\sum\limits_{SR}^{}(\lambda_{SPF} + \lambda_{RF})}{\sum\limits_{SR}^{}\lambda}$$

where $\lambda_SPF$ is the failure rate associated with hardware element single-point faults, $\lambda_RF$ is the failure rate associated with hardware element residual faults, and $\lambda$ is the failure rate associated too all faults. 

\begin{figure}[!h]
    \centering
	\includegraphics[width=1\linewidth]{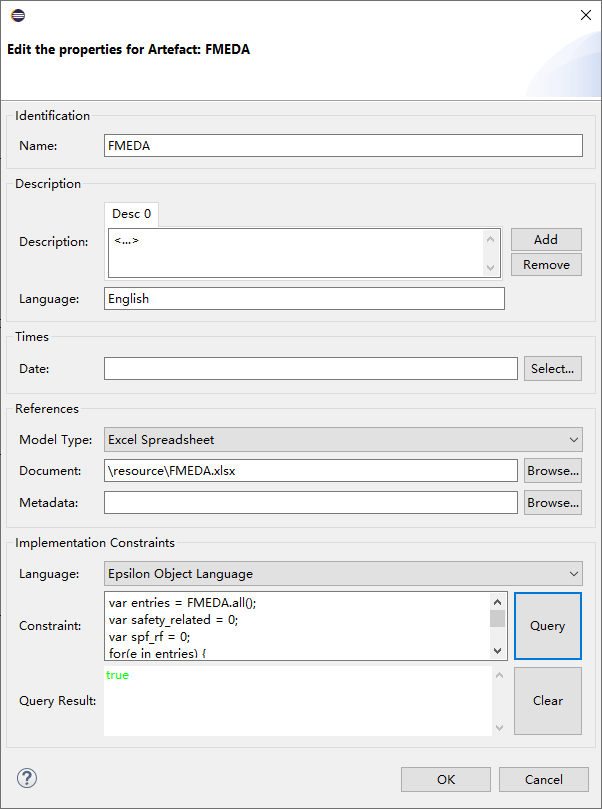}
	\caption{Reference to FMEDA in Excel spreadsheet.}
	\label{fig:fmeda}
\end{figure}

\begin{lstlisting}[language=EOL, caption={Computing the SPFM for a hardware component.},label={lst:spfm}]
var entries = FMEDA.all();
var safety_related = 0;
var spf_rf = 0;
for(e in entries) {
	if(e.SafetyRelated = "Yes") {
		safety_related += e.FailureRate.asReal();
	}
	if(e.SafetyGoalViolation = "Yes") {
		spf_rf += e.SPF_RF.asReal();
	}
}
var spfm = 1 - (spf_rf)/safety_related;
return spfm > 0.9;
\end{lstlisting}

\begin{figure}[!h]
    \centering
	\includegraphics[width=1\linewidth]{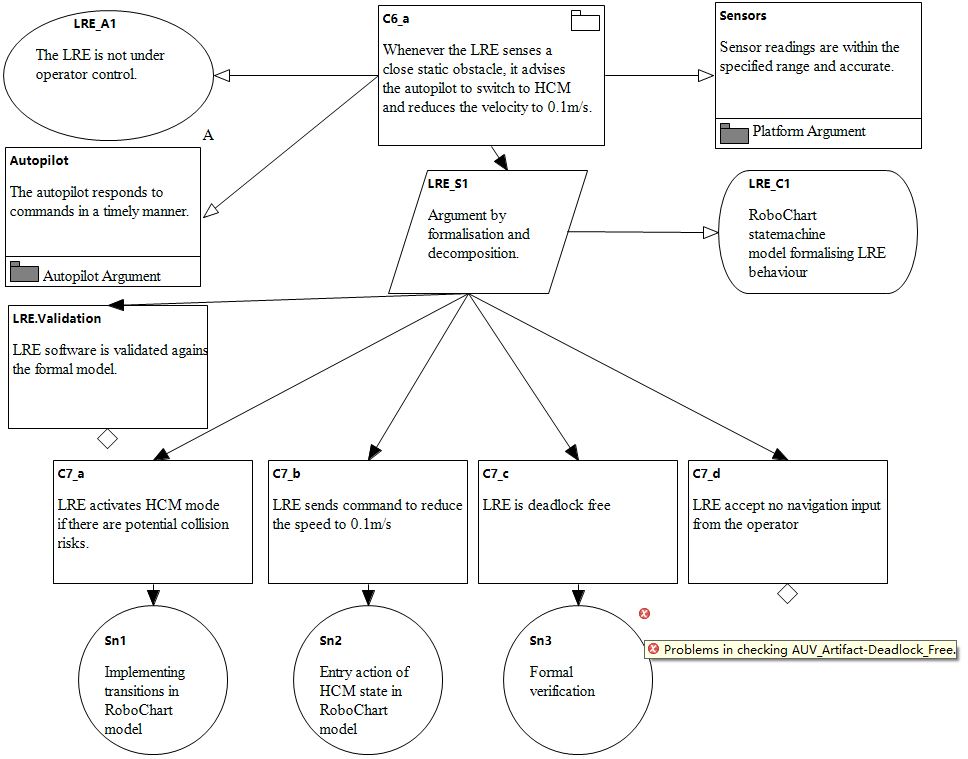}
	\caption{Fragment in the LRE assurance case module to argue the safety of static obstacle avoidance.}
	\label{fig:auv_GSN_module}
\end{figure}

To validate the FMEDA, we aim to achieve the target SPFM (assume we aim to achieve 90\%). 
In order to do this automatically in ACME, in the assurance case, we create an \textbf{ArtifactPackage} (discussed in Section~\ref{sec_sacm}), and in it create an \textbf{Artifact}, and we refer to the Excel spreadsheet from the \textbf{Artifact}, which can be later used as evidence to our safety argument in the assurance case. 
As shown in Figure~\ref{fig:fmeda}, in the \textit{references} section, we configure the ``Model Type'' (as Excel spreadsheet), the ``Docuemtn'' (to the location of the Excel spreadsheet), and the ``Metadata'' (empty in this example). 
Within ACME, we use \textbf{ImplementationConstraint}s to store model validation rules in \textbf{Artifact} elements so that such rules can be executed when we evaluate the assurance case.
In Figure~\ref{fig:fmeda}, in the ``Constraint'' section, we use the rule in Listing~\ref{lst:spfm} (written in the Epsilon Object Language -- EOL \cite{kolovos2006epsilon}) to check if the FMEDA fulfils our requirement for the target SPFM value.
During the development process, when the FMEDA changes, ACME can automatically detect this change by the execution of the query and compute the SPFM automatically, then show the users if the hardware design fails to meet the target SPFM values.

With the safety analysis performed, we are able to further refine our assurance case by arguing the safety of subsystems (task \textbf{A4}), a fragment of the assurance case is shown in Figure~\ref{fig:auv_GSN_module}.
However, we shall note that in this ACCESS step, only \textit{C6\_a} and \textit{Autopilot} in Figure~\ref{fig:auv_GSN_module} are developed (other elements are defined in the next ACCESS step).
Also, at this stage, there are no traceability from the argument to the supporting evidence yet.

\subsection{ACCESS Step 4}
\begin{figure*}[h!]
    \centering
	\includegraphics[width=1\linewidth]{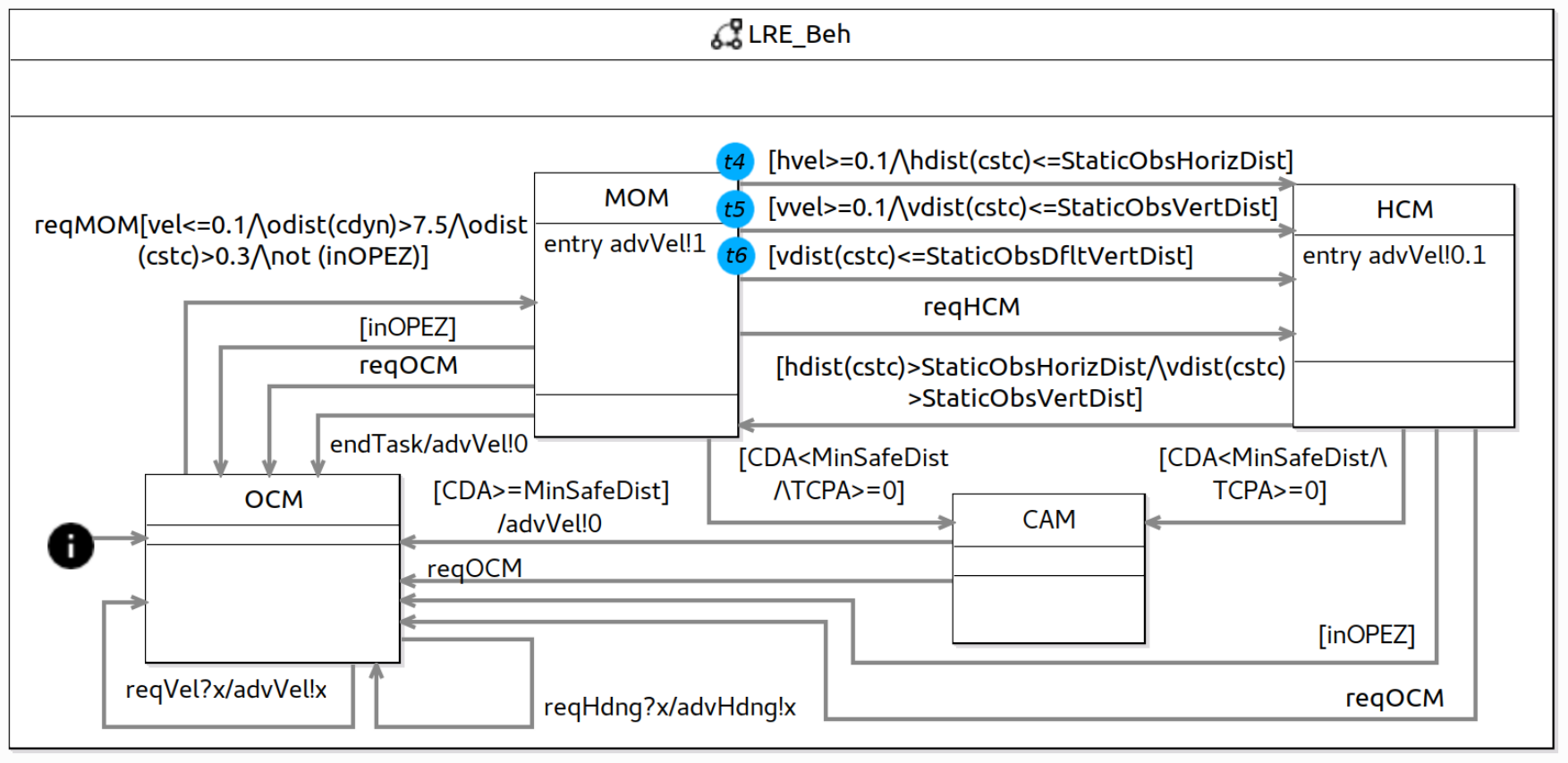}
	\caption{LRE RoboChart State Machine}
	\label{fig:lre_statemachine}
\end{figure*}
\subsubsection{The Last Response Engine}
In this paper, we focus on the development of the LRE, which provides run-time safety assurance. 
We consider the case of the AUV navigating within an enclosed pond to perform maintenance tasks. 
There are two main hazards for the AUV that we consider: (1) collisions with static and dynamic obstacles and (2) causing a splash, which can be a hazard for workers and equipment around the pond. 
The AUV can either be under operator control, or running autonomously. 
If operating autonomously, the responsibility for satisfying the safety requirements lies with the LRE, which can engage evasive maneuvers if necessary. 
There are also \textit{Object Proximity Exclusion Zone}s (OPEZs), which are designated areas where the AUV may not operate autonomously, and help with hazard mitigation. 
They include the area close to the pond wall, and also the area just below the water surface.

The LRE functions in four modes: Operator Control Mode (OCM), Main Operating Mode (MOM), High Caution Mode (HCM) and Collision Avoidance Mode (CAM).
In OCM, the LRE passes control inputs from the operator to the autopilot.
In MOM, the LRE takes control for normal behaviour at maximum speed.
HCM is for the situation when the AUV is getting close to an obstacle, and so the LRE lowers the velocity.
Finally, CAM is the mode where a potential unavoidable collision has been detected, and the AUV is manoeuvring away from the obstacle.

The LRE keeps an \textit{obstacle register}, which stores identified obstacles, through sensor readings.
In each behavioural cycle, the LRE calculates the closest obstacle and determines whether it should apply evasive manoeuvres or switch into high caution mode (HCM).

There are six event inputs: \textit{reqVel}, with which the operator can request a velocity; 
\textit{reqHdng}, to request a new heading; 
\textit{reqOCM}, \textit{reqMOM}, \textit{reqHCM}, to request an operation mode; 
and \textit{endTask}, to delineate tasks.
The two output events are \textit{advVel} and \textit{advHdng}, with which the LRE can send instructions to change velocity or heading to the autopilot.

\subsubsection{Behaviour model for the LRE}
We now model the behaviour of the LRE (task \textbf{S5}).
We create a state machine for the LRE, shown in Figure~\ref{fig:lre_statemachine}.
It implements the LRE's behavioural requirements and specifies the conditions on switching to different operation modes.
The following definitions and functions appear in the state machine~\cite{miyazawa2016robochart}:
\textit{vel} (velocity of the AUV), \textit{inOPEZ} (if the AUV is in an OPEZ), \textit{CDA} (Closest Distance of Approach), \textit{StaticObsHorizDist} and \textit{StaticObsVertDist} (shortest distance allowed to an obstacle horizontally and vertically), \textit{MinSafeDist} (minimal overall safe distance), \textit{cdyn} (closest dynamic obstacle), \textit{cstc} (closest static obstacle), \textit{hdist()} (horizontal distance to an obstacle), \textit{vdist()} (vertical distance to an obstacle), \textit{odist()} (overall distance to an obstacle).

The transitions give (1) events that trigger the transition; (2) the conditions under which they can fire, and (3) any action taken at that point. 
For example, the top-left most transition in Figure~\ref{fig:lre_statemachine} is

$$reqMOM\left[\begin{array}{l} vel \le 0.1 \land odist(cdyn) > 7.5 \\
          \land odist(cstc) > 0.3 \land \neg \textit{inOPEZ}
        \end{array}\right]$$

It states that the LRE can move from OCM to MOM when the trigger event $reqMOM$ is received from the operator, and the
set of conjoined conditions hold.
Specifically, the AUV can only operate autonomously provided it has a low velocity, a minimum distance to static and dynamic obstacles, and the AUV is not in an OPEZ.
The state $MOM$ has an entry action, $advVel!1$, that is executed when the state is activated from any transition, and advises the autopilot to set the velocity to the maximum.
The top-most transition from MOM to HCM has no trigger action, and only the condition $$\left[hvel \ge 0.1 \land hdist(cstc) \le \textit{StaticObsHorizDist}\right]$$ attached, meaning that it will activate as soon as the sensor values enter the characterised range.

\subsubsection{LRE Argumentation}
With the behavioural model for the AUV defined, we discuss model analysis and verification (task \textbf{A5}) and further refined the assurance case (created using ACME) for the LRE,.

We focus on the scenario of static obstacle avoidance for the LRE, the safety argument fragment of which is shown in Figure~\ref{fig:auv_GSN_module}.
The top level \textbf{Goal} \textit{C6\_a} states that upon detecting a close static obstacle, LRE should advise the autopilot to switch to HCM and reduce the velocity of the AUV to 0.1m/s.
\textit{C6\_a} is a public goal (indicated by the module icon on the top right corner) as it is used by the overall safety argument in the \textit{AUV System} module, unlike other goals, which are private.
\textit{C6\_a} is in the context of, and thus contingent upon, \textbf{Assumption} \textit{LRE\_A1}, and \textbf{Away Goal}s \textit{Autopilot} and \textit{Sensors}. The away goals must be supported in the \textit{Platform} and \textit{Autopilot} modules for the LRE module to be valid. \textit{LRE\_A1} ensures that the argument need only hold when the operator is not in control; the alternative case is handled by the \textit{Operator} module.
We support \textit{C6\_a} by formalisation and decomposition. 
\textbf{Strategy} \textit{LRE\_S1} states our argument strategy, which is in the context of \textbf{Context} \textit{LRE\_C1}.

We focus on \textbf{Goal}s \textit{C7\_a}, \textit{C7\_b} and \textit{C7\_c}. They use the RoboChart model to establish that the safety requirement is indeed satisfied. They are subject to a validation argument under \textit{LRE.Validation}, which is left undeveloped for now, but should include activities like software testing.
In \textit{C7\_a}, we state that the LRE should activate HCM if there are potential collision risks. We support this \textbf{Goal} with \textit{Solution} \textit{Sn1}, which states that transitions to HCM mode from MOM should be modelled by the behavioural model in Figure~\ref{fig:lre_statemachine}.
In \textit{C7\_b}, we state that the LRE should send a command to the autopilot to reduce the speed to 0.1m/s, and we support this with \textbf{Solution} \textit{Sn2}, which states that the entry action of HCM should reduce the speed to 0.1m/s.
In \textit{C7\_c}, we state that the LRE is deadlock free, and support this with \textbf{Solution} \textit{Sn3} by formal verification\footnote{We will explain the error marker on Sn3 later}.

At this stage, we achieve traceability to formal verifications inside the assurance case.
But the assurance case is by no means complete, since systematic verification and validation are yet to be performed.

\begin{figure}[!h]
    \centering
	\includegraphics[width=1\linewidth]{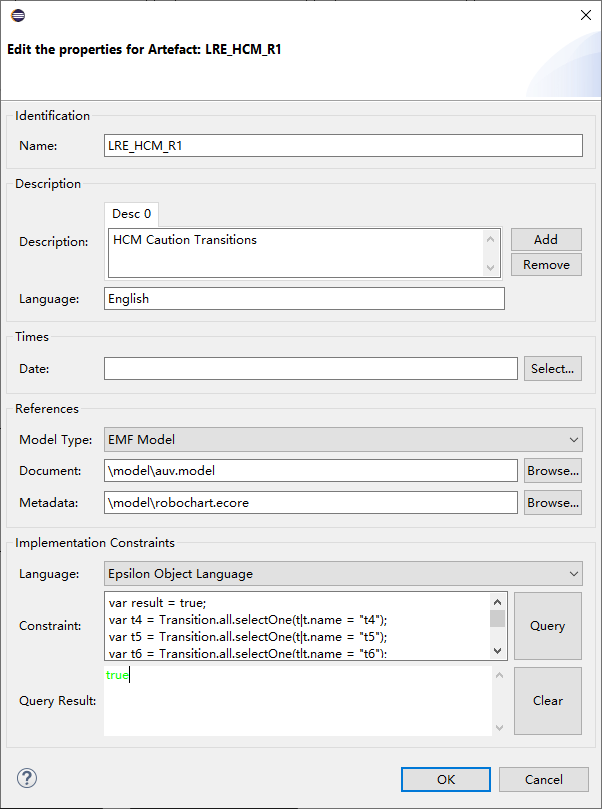}
	\caption{ACME dialog to edit \textbf{Artifact} \textit{LRE\_HCM\_R1}.}
	\label{fig:artifact_dialog}
\end{figure}

\subsection{ACCESS Step 5}
In this step, the system shall be implemented (task \textbf{S6}), and we perform system verification and validation (task \textbf{A6}) from assurance case.
We do so by complete the traceability to all engineering artifacts from the assurance case, and automate the evaluation from ACME.
\subsubsection{Trace to EMF models}
\label{sec:trace_emf}
GSN elements such as \textbf{Context}s and \textbf{Solution}s, can refer to models/documents external to the assurance case. 
With traditional GSN approaches, references to external models/documents are informal and their evaluation is often performed manually.

We illustrate traceability with \textbf{Goal} \textit{C7\_a} and its supporting \textbf{Solution} \textit{Sn1} (in Figure~\ref{fig:auv_GSN_module}), which in turn is supported by several transitions in the RoboChart state machine.
To be able to reference elements of the RoboChart model shown in Figure~\ref{fig:lre_statemachine}, we create an \textbf{Artifact} named \textit{LRE\_HCM\_R1} (in an \textbf{ArtifactPackage} named \textit{LRE\_Artifact}), which will be referenced by \textbf{Solution} \textit{Sn1}. 
The properties of \textit{LRE\_HCM\_R1} are shown in Figure~\ref{fig:artifact_dialog}.
In the properties view for an \textbf{Artifact}, we specify the ``Model Type'' (we currently support EMF models, Excel spreadsheets and plain text files), ``Document'' (location of the model) and the ``Metadata'' (metadata of the document, which can be metamodels, schemas, etc.), in the ``References'' section in Figure~\ref{fig:artifact_dialog} (Note that assurance cases and its referenced engineering models should reside in the same location).

We then attach the model validation rule in Listing~\ref{lst:query} in the ``Constraint'' section.
In this rule we check that there are at least 3 transitions from MOM, named ``t4'', ``t5'' and ``t6'' (shown in Figure~\ref{fig:lre_statemachine}), which are triggered when there are potential collision risks.
For readability we only show the queries for Transition ``t4''.
The user can evaluate the query inside the dialog by pressing the ``Query'' button. 
ACME will load the model specified in the ``Reference'' section and execute the query, the result of which is displayed in the ``Query Result'' text field.
It is to be noted that the validation rules do not have to be specified by EOL, in a separate publication~\cite{wei2023automated}, we illustrated the use of constraint natural language with model-based approach. 

\begin{lstlisting}[language=EOL, caption={Query the transitions in the RoboChart model.},label={lst:query}]
var result = true;
var t4 = Transition.all.selectOne(t|t.name = "t4");
var t5 = Transition.all.selectOne(t|t.name = "t6");
var t6 = Transition.all.selectOne(t|t.name = "t7");
var t4c = t4.condition;
var t4check = t4c.isTypeOf(And) and
t4c.left.isTypeOf(GreaterOrEqual) and
t4c.left.left.ref.name = "hvel" and
t4c.left.right.value = 0.1 and
t4c.right.isTypeOf(LessOrEqual) and
t4c.right.left.isTypeOf(CallExp) and
t4c.right.left.function.ref.name = "hdst" and
t4c.right.left.args.first.ref.name = "cstc" and
t4c.right.right.ref.name = "StaticObsHorizDist";
result = result and t4check;
return result;
\end{lstlisting}

\begin{figure}[!h]
    \centering
	\includegraphics[width=1\linewidth]{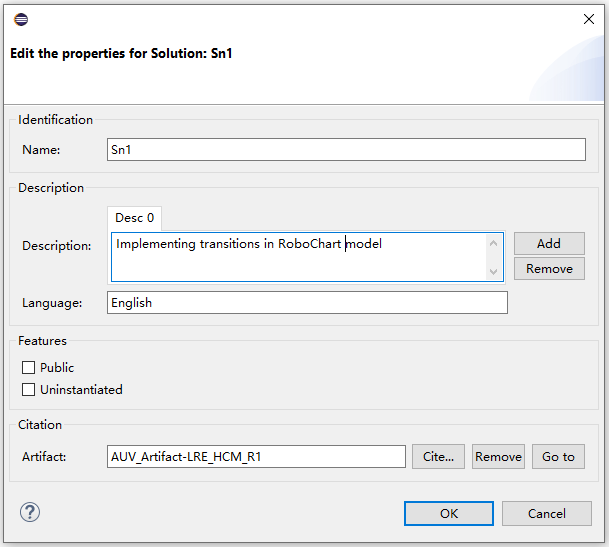}
	\caption{ACME dialog to edit \textbf{Solution} \textit{Sn1}.}
	\label{fig:solution_citation}
\end{figure}

\textbf{Artifact} \textit{LRE\_HCM\_R1} can then be used as a supporting evidence for our assurance case (specifically to substantiate \textbf{Solution} \textit{Sn1}). 
To do this, within \textit{Sn1} we ``cite'' the \textit{LRE\_HCM\_R1} (defined in the \textit{AUV\_Artifact} package) in the ``Citation'' section, shown in Figure~\ref{fig:solution_citation}.

\subsubsection{Trace to Isabelle Theory Files}
\begin{figure}[!h]
    \centering
	\includegraphics[width=0.8\linewidth]{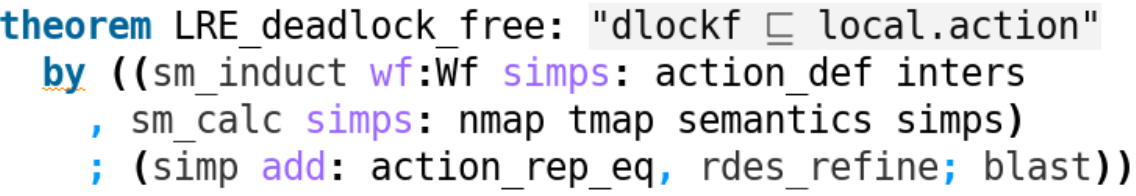}
	\caption{Deadlock free theorem defined in Isabelle.}
	\label{fig:deadlock_free}
\end{figure}

In this work, we also support references to formal notations embedded in Isabelle theory files.
The process of referring to an Isabelle theory file is the same described in Section~\ref{sec:trace_emf}, except that the users need to select ``Isabelle Theory File'' in the ``Model Type'' drop down menu (within the ``References'' section) in the property dialog of an \textbf{Artifact}, so that ACME knows to invoke the Isabelle server to check the referenced theory file.
Upon assurance case evaluation, ACME sends the Isabelle theory file (\textit{Formalisation}) to the \textit{Isabelle Server}, which checks the theory file and returns a JSON string, ACME then parses the string and marks the errors in the \textbf{ArtifactPackage} editor.
We illustrates this by injecting an error and shows this in ACME, as shown in lower part of Figure~\ref{fig:approach}. 
For this work, we create an \textbf{Artifact} called \textit{Deadlock\_Free} and refer to the theorem shown in Figure~\ref{fig:deadlock_free}. 
This uses automated proof tactics in Isabelle to prove deadlock freedom for the LRE state machine in Figure~\ref{fig:lre_statemachine}.
We then ``cite'' \textit{Deadlock\_Free} within \textbf{Solution} \textit{Sn3} in Figure~\ref{fig:auv_GSN_module}.

\subsubsection{Automated Assurance Case Evaluation}
With engineering artifacts referenced from our assurance case, we can evaluate the assurance case by invoking the ``evaluate'' function, which users can select in the context menu provided by ACME.
When we evaluate the assurance case in ACME, ACME starts the evaluation process from the assurance case diagram where the ``evaluate'' function is invoked.
ACME automatically traces to \textbf{Artifact}s from \textbf{Solution}s, \textbf{Context}s, and \textbf{Assumptions}. 
Then, depending on the types of the engineering artifacts (this information is associated to the \textbf{Artifact}s), ACME executes model queries (for model validation) or invokes the Isabelle server (for formal verification), respectively.
Figure~\ref{fig:auv_GSN_module} shows an error marker on \textbf{Solution} \textit{Sn3}, which indicates that the Isabelle proof in Figure~\ref{fig:deadlock_free} referenced from \textbf{Artifact} \textit{Deadlock\_Free} was unsuccessful.

This process of evaluation can be performed at regular intervals to ensure that updates to models and other artifacts do not invalidate the assurance case. 
For example, if one of the transitions from MOM to OCM is removed in the behavioural state machine, ACME will be able to pick up this change and flag an error. 
Moreover, if a change to the state machine introduces deadlock, this will also be flagged by the failure of the proof in Figure~\ref{fig:lre_statemachine}. 
This is typically the benefits by adopting ACCESS, ACCESS assumes that model-based approaches are used in the development process, hence boosting the efficiency of developers and improving the correctness and coverage of assurance case evaluation activities.

\subsubsection{Transformation to Isabelle/SACM}
Once a satisfactory assurance case is developed, a further step is to check the integrity of the overall assurance case formally using Isabelle/SACM.
This is an optional step for ACCESS, but we would like to show that how formalism can be integrated with model-based assurance case to form a more convincing assurance case.
To do this, we perform a model-to-text transformation to generate Isabelle/SACM notations from the assurance case automatically, which can be machine-check using the Isabelle server.
In ACME, we use the Epsilon Generation Language (EGL)~\cite{rose2008epsilon} to implement the transformation, but the transformation can be generalised.
Algorithm~\ref{alg:a1} shows the pseudo-code for generating Isabelle/SACM notations from a GSN module.
\begin{algorithm}[h!]
	{
		\fontsize{9}{10}
		\selectfont
		\SetAlgoLined\DontPrintSemicolon
		\For {element in \{all Contextual Elements\} $\cup$ \{all Goals\}} {
		    \Let declarations = \{``'', ``axiomatic'', ``assumed'', ``needsSupport'', ``asserted''\};\\
		    \Let declaration = determined based on the feature/type of element\\
    		\Output{``Claim '' + \textit{element.name} + \textit{declaration} + ``\guilsinglleft'' + \textit{element.description} + ``\guilsinglright''}\\
		}
		\For {element in \{all Solutions\}} {
		\Output{``ArtifactReference'' + \textit{element.name} + \textit{declaration} + ``\guilsinglleft'' + \textit{element.description} + ``\guilsinglright''}\\
		}
		\For {element in \{all Strategies\}} {
		    \Let target = incoming SupportedBy\\
		    \Let sources = outgoing SupportedBys\\
		    \Let source\_names = ``@\{Claim'' + \textit{source names separated by ``,''} +``\}''\\
		\Output{``Inference '' + \textit{element.name} + `` src \guilsinglleft\{'' + \textit{source\_names} + ``\}\guilsinglright ~tgt \guilsinglleft\{@\{Claim '' + \textit{target.name} + ``\}\}\guilsinglright'' + `` \guilsinglleft@\{Claim '' + \textit{target.name} + ``\} is supported by '' + \textit{source\_names} + ``.\guilsinglright'';}\\
		}
		\For {element in \{all Relationships Not Processed\}} {
		\Let source\_name = element.target.name;\\
		\Let target\_name = element.source.name;\\
		\If{element isTypeOf(GSN!SupportedBy)}{
		    \lIf{source.isTypeOf(GSN!Solution)}{
		    \Output{``Inference '' + \textit{element.name} + ``src \guilsinglleft\{@\{ArtifactReference '' + \textit{source\_name} + ``\}\}\guilsinglright ~tgt \guilsinglleft\{@\{Claim'' +  \textit{target\_name} +"\}\}\guilsinglright" + `` \guilsinglleft@\{Claim '' + \textit{target\_name} + `` \} is supported by @\{ArtifactReference '' + \textit{source\_name} + ``\}.\guilsinglright''}
		    }
		    \lElse{\\
		    \Output{``Inference '' + \textit{element.name} + ``src \guilsinglleft\{@\{Claim '' + \textit{source\_name} + ``\}\}\guilsinglright ~tgt \guilsinglleft\{@\{Claim'' +  \textit{target\_name} +``\}\}\guilsinglright'' + `` \guilsinglleft@\{Claim '' + \textit{target\_name} + `` \} is supported by @\{ArtifactReference '' + \textit{source\_name} + ``\}.\guilsinglright''}
		    }
	    	}
	    	\ElseIf{element isTypeOf(GSN!InContextOf)}{
		       \Output{``Context '' + \textit{element.name} + ``src \guilsinglleft\{@\{Claim '' + \textit{source\_name} + ``\}\}\guilsinglright ~tgt \guilsinglleft\{@\{Claim'' +  \textit{target\_name} + ``\}\}\guilsinglright'' + ``\guilsinglleft@\{Claim '' + \textit{target\_name} + `` \} is context for @\{Claim '' + \textit{source\_name} + ``\}.\guilsinglright''}
	    	}
	    }
	    }
	\caption{Generating Isabelle from GSN.}
	\label{alg:a1}
\end{algorithm}

\begin{figure}[!h]
    \centering
    \includegraphics[width=0.8\linewidth]{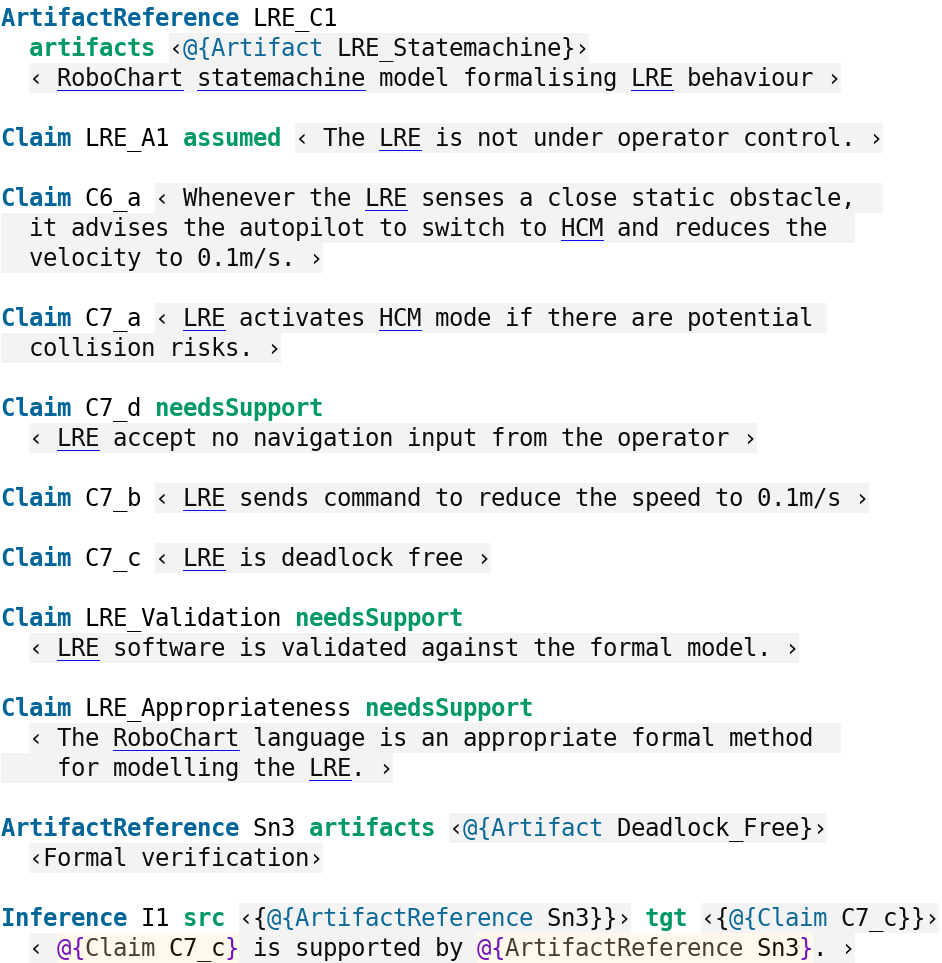}
    \caption{LRE Argument Fragment in Isabelle/SACM}
    \label{fig:lre_isa_sacm}
\end{figure}

A fragment of transformed Isabelle/SACM argument is shown in Figure~\ref{fig:lre_isa_sacm}, with some elements reordered to aid readability. 
\textbf{Context} \textit{LRE\_C1} and \textbf{Solution} \textit{Sn3} both give rise to references to formal artifacts included from another Isabelle theory file~\cite{foster2020formal}, including the formalised LRE state machine. 
Several of the claims are left open, and so have the keyword \textit{needsSupport} attached, which can be checked to ensure that all branches of the argument are developed. 
An \textbf{Inference}, \textit{I1}, connects the formalised \textit{Sn3} (the source) to the \textbf{Claim} \textit{C7\_c} (the target). 
The remainder of the claims, inferences, and solutions are similarly mapped.

The transformed Isabelle/SACM theory file is sent to the remote Isabelle server for machine-checking.
Inside ACME we parse the JSON string returned by Isabelle and find out if there are any problems. If so we can locate model-elements in the assurance case that cause problems and display them in ACME editors.

\subsection{ACCESS Step 6}
\subsubsection{Dynamic Assurance Case Evaluation for AUV at Runtime}
As previously discussed, the assurance of open adaptive RAS requires safety evaluation to be performed at runtime when the system is operational.
To take a first practical step in this direction, we show how we can achieve dynamic assurance case evaluation with dynamic runtime data, in order to support the ACCESS process.

In our LRE assurance case in Figure~\ref{fig:auv_GSN_module}, we state that \textbf{Goal} \textit{C6\_a} is valid within the context of \textbf{Away Goal} \textit{Sensors} (top right element in Figure~\ref{fig:auv_GSN_module}), which is a top level \textbf{Goal} specified in the \textit{Platform\_Argument} module (hence the SupportedBy relation in Figure~\ref{fig:auv_overall}).

\begin{figure}[!h]
    \centering
	\includegraphics[width=0.8\linewidth]{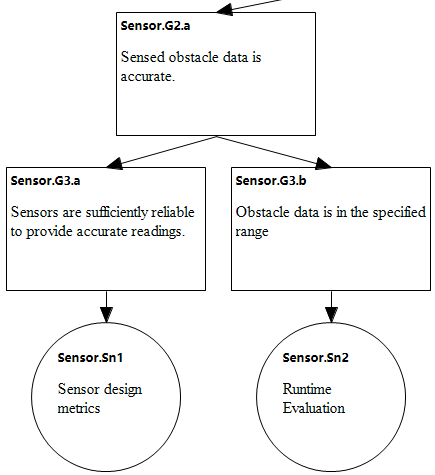}
	\caption{\textbf{Goal} in the \textit{Sensor} \textbf{Module} that requires runtime evaluation.}
	\label{fig:dsms}
\end{figure}

In module \textit{Platform\_Argument}, we define a \textbf{Goal} \textit{Sensor.G2.a} shown in Figure~\ref{fig:dsms} (we omit other parts of the argument in the \textit{Platform\_Argument} module to aid readability), which supports the top level \textbf{Goal} in the \textit{Platform\_Argument} module. 
In turn, two more \textbf{Goal}s that support \textit{Sensor.G2.a}: \textbf{Goal} \textit{Sensor.G3.a} states that sensors should be sufficiently reliable to provide accurate readings, this is in turn supported by \textbf{Solution} \textit{Sensor.Sn1}, where the hardware design metrics is quantitatively analysed by FMEDA (we omit the details of the analysis for readability); \textbf{Goal} \textit{Sensor.G3.b} states that obstacle data should be in the specified range, which is supported by \textbf{Solution} \textit{Sensor.Sn2}, where runtime evaluation is performed.

This is where the assurance case starts to deviate into a dynamic assurance case. 
Conventionally, assurance cases have static evidence in the argument structure, which means, \textbf{Goal} \textit{Sensor.G2.a} needs only to be supported by \textbf{Goal} \textit{Sensor.G3.a}, because there was no notion of detecting random hardware faults at runtime.
In our approach, we argue that runtime random hardware fault can be detected by means of evaluating dynamic assurance cases.
In this way, we can establish our confidence in the reliability of the entire system, due to the fact we could establish our confidence in the reliability of key components of the system (rather than trusting their manufacture specifications on failure rates). 
Therefore, \textbf{Solution} \textit{Sensor.Sn2} is in place for ACME runtime component to perform dynamic validation (this corresponds to task \textbf{A7}).

\begin{figure}[!h]
    \centering
	\includegraphics[width=.6\linewidth]{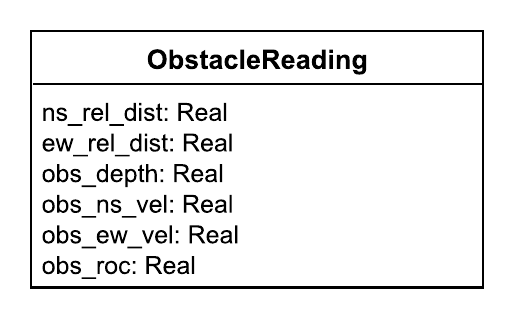}
	\caption{ObstacleReading in the LRE Runtime Assurance Metamodel.}
	\label{fig:dt_or}
\end{figure}

\begin{figure}[!h]
    \centering
	\includegraphics[width=.7\linewidth]{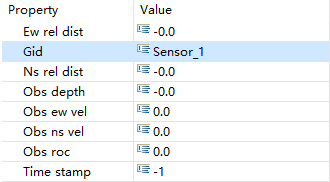}
	\caption{Example \textit{ObstacleReading} in the sensor digital twin model.}
	\label{fig:sensor_model}
\end{figure}

To determine the reliability of the sensors , we need to synchronise their readings to models that we can refer to at runtime from our assurance case.
For this purpose, we create an \textit{LRE Runtime Assurance} metamodel.
We are particularly interested in the \textit{ObstacleReading} class in the metamodel, which is shown in Figure~\ref{fig:dt_or}.
We are interested in 6 variables in an \textit{ObstacleReading}, \textit{ns\_rel\_dis} for relative north-east distance to the obstacle, \textit{ew\_rel\_dist} for relative east-west distance to the obstacle, \textit{obs\_depth} for the depth of the obstacle, \textit{obs\_ns\_vel} for north-east velocity of the obstacle, \textit{obs\_ew\_vel} for east-west velocity of the obstacle, and then \textit{obs\_roc} for rate of climb of the obstacle.
With the \textit{LRE Runtime Assurance} metamodel, we create an \textit{LRE Runtime Assurance} model, which contains default content before the AUV is deployed, shown in Figure~\ref{fig:sensor_model}.

In our \textbf{ArtifactPackage}, we create an \textbf{Artifact} \textit{Obstacle\_reading}, which refers to the \textit{LRE Runtime Assurance} model.
Inside \textit{Obstacle\_reading}, we attach the validation rule in Listing~\ref{lst:check_readings}, which checks that the sensor readings are within the valid ranges for the AUV.
By performing the above activities, we have converted some elements within the model-based assurance case into runtime elements, the validity of which would be determined by runtime data.

\begin{lstlisting}[language=EOL, caption={Evaluation rule for checking the well-formedness of obstacle readings.},label={lst:check_readings}] 
var r = M!ObstacleReading.all().first();
return (r.ns_rel_dist>=-50.0 and r.ns_rel_dist<=50.0)
and (r.ew_rel_dist>=-50.0 and r.ew_rel_dist<=50.0)
and (r.obs_depth>=-10.0 and r.obs_depth<=0.0)
and (r.obs_ns_vel>=-5.0 and r.obs_ns_vel<=5.0)
and (r.obs_ew_vel>=-5.0 and r.obs_ew_vel<=5.0)
and (r.obs_roc>=-5.0 and r.obs_roc<=5.0);
\end{lstlisting}

\subsection{ACCESS Step 7}
For each class in the \textit{LRE Runtime Assurance} metamodel, we use model-to-text transformation to generate a Java class \textit{XXXDriver} (details omitted to aid readability).
For example, for the \textit{ObstacleReading} class, we generate \textit{ObstacleReadingDriver}. 
To synchronise the sensor data, we create a Java class \textit{ObstacleReadingDriver} which is a type of \textit{Runtime Data Driver}, as discussed in Figure~\ref{fig:approach}.
\textit{ObstacleReadingDriver} provides an API which takes 6 parameters (\textit{ns\_rel\_dist}, \textit{ew\_rel\_dist}, \textit{obs\_depth}, \textit{obs\_ns\_vel}, \textit{obs\_ew\_vel} and \textit{obs\_roc}), for the AUV to write these values to the \textit{AUV Sensor} model at runtime.
For runtime evaluation, ACME provides a \textbf{Dynamic Safety Management System} (DSMS), as shown in Figure~\ref{fig:approach}.

In terms of model evaluation, we actively evaluate the assurance case by executing the ``evaluate'' function of the DSMS, which in turn executes all model validation rules embedded in the \textbf{Artifact}s (which support \textbf{Solution}s, \textbf{Context}s, \textbf{Assumption}s and \textbf{Justification}s) in the assurance case.
In our prototype, the DSMS periodically evaluates the AUV assurance case (in this work we use a 50-millisecond intervals), which includes the evaluation of the \textbf{Solution} \textit{Senso.G3.b}, which then triggers the evaluation of the \textbf{Artifact} \textit{Obstacle\_reading} by executing the evaluation rule in Listing~\ref{lst:check_readings} against the \textit{AUV Sensor} model.

The runtime data is synchronised from the AUV to the \textit{LRE Runtime Assurance} model through \textit{ObstacleReadingDriver} (through simulation, as the AUV has not been developed for operational trials). 
If the readings are not within the range defined in the evaluation rule, the periodic evaluation of DSMS will fail, rendering the corresponding assurance case fragment invalid.
In a sense, the development time assurance case is converted to a \textit{model@runtime} with the DSMS framework.
In this way, a novel means to perform additional checks at runtime (and reflect back to the assurance case) is proposed. 
A number of advantages comes with this approach: 
\begin{itemize}
    \item Random hardware faults is actively detected rather than analysed and calculated, this may result in a shorter \textit{Time to Detect Fault}; in addition, this fault is propagated back to the assurance case, rendering it invalid immediately. At runtime, the system may consult DSMS, if the assurance case becomes invalid, it may make the transition to safe state (e.g. complete hault), result in shorter \textit{Fault Reaction Time Interval};
    \item The \textit{LRE Runtime Assurance} model can be version-controlled, which makes it possible to accumulate history data for the corresponding sensor, therefore obtaining  statistics on failure rates for the sensor, which can be further used in stochastic approaches approximating the fault rate of the sensors. 
    \item Consequently, the safety analysis for the sensors can be updated accordingly, which we discussed in a recently published work~\cite{wei2022designing, wei2023decisive}.
\end{itemize}

It is to be noted that \textbf{Context}s in the assurance case may also depend on sensor readings.
For RAS, they may operate in different contexts, the safety arguments for such contexts may differ. 
Therefore, it is highly likely that dynamic assurance cases for RAS are no longer singular assurance case, but a repository of models, including a number of assurance cases for different operational contexts, engineering artifacts produced at development time, as well as \textit{models@runtime} with synchronised sensor data. 
The repository of models should be monitored and version-controlled at runtime to aid automated assurance case evaluation and offline retrospection. 
To this point, we have concluded \textbf{ACCESS Step 7}. 
It is to be noted that ACCESS is an iterative process, development following ACCESS can start any ACCESS step in the development lifecycle, based on the justifications of the developers.

%% file: section_6.tex
\section{Evaluation}
\label{sec:eval}
To evaluate the effectiveness and generality of ACCESS, we used our methodology to engineer the UAV (hereby reffered to as System A) from Section~\ref{sec:case}, and a safety-critical autonomous robotic system (hereby reffered to as System B) for earthquake aftermath search and rescue.
The two systems were developed as described in Section~\ref{sec:access}, and were deployed in a realistic environment with changes in environment specific to their application domains. 
We examined the efficiency of the development process for both systems, and evaluate the generality of our methodology across differen domains. In addition, we examine the coverage and the scalability of the supporting too, which provides insights on future research directions based on ACCESS and any model-based assurance case management environment.
The aim of our evaluation was to answer the following research questions. 

\vspace{2mm}
\textbf{RQ1} (Efficiency): Does ACCESS, supported by model-based assurance case, increase efficiency of developers for safety critical systems?

\vspace{2mm}
\textbf{RQ2} (Generality): Does ACCESS support the development of safety critical systems and model-based assurance cases across application domains?

\vspace{2mm}
Although the focus of research contribution is on the proposed ACCESS methodology, we also aim to answer the following research questions.

\vspace{2mm}
\textbf{RQ3} (Coverage): Does ACCESS supported with ACME, cover the types of heterogeneous models that are produced throughout the development of safety critical systems to achieve the highest degree of automation?

\vspace{2mm}
\textbf{RQ4} (Scalability): Does ACCESS supported with ACME, support the development of complex safety critical systems with a large number of model elements?

\vspace{2mm}
As the focus of our evaluation is the ACCESS methodology (primary) and its tool support (secondary), we necessarily made a number of assumptions. 
In particular, we assumed that the development activities identified in the ACCCESS methodology are (mostly) model based, which produce structured models that can be processed by model management frameworks. Secondly, we assumed that ACCESS could be used to construct assurance cases for all aspects of the target system, including their design, development, operation and maintenance. We further assumed that the runtime assurance case converted from development time assurance case, would be used by the target system as an non-invasive service (i.e. the runtime assurance case does not interact directly with the decision-making components of the target system), the target system would consult the runtime evaluation results of the assurance case and adapt accordingly.

The experiments carried out to address the above research questions are described in Sections \ref{sec:efficiency}, \ref{sec:generality}, \ref{sec:coverage} and \ref{sec:scalability}.

\subsection{RQ1: Efficiency}
\label{sec:efficiency}
For efficiency, we conducted comparative experiments on sub-components of System A and System B due do their complexities.
The evaluation set-up is as follows; 

\begin{enumerate}
    \item Evaluation subjects. For System A, we selected a power supply unit (Subsystem A) as the experiment subject. For System B, we selected a navigation unit (Subsystem B) which receives sensor data and plans the route for the system.
    \item Participants. We asked three safety-critical systems engineer (with relatively same level of expertise) to participate in the experiments. 
\end{enumerate}

The experiment procedure is as follows. We asked Participant A to engineer Subsystem A and Subsystem B by following ACCESS with a complete manual approach with no model-based tool support (i.e. all artifacts produced are text-based documents). 
We then asked Participant B to engineer Subsystem A and Subsystem B (by following ACCESS) with an model-based approach but without the support from ACME, Participant B chose to use EMF~\cite{steinberg2008emf} and Robochart~\cite{Miyazawa2019-RoboChart} as the modeling platform and chose Eclipse Epsilon~\cite{kolovos2008epsilon} as the model management framework. 
We then asked Participant C to engineer Subsystem A and Subsystem B (by following ACCESS) with the tool support from ACME. 
We recorded the effort it took for all three participants to engineer Subsystem A and Subsystem B in terms of time.
The experiment results are shown in Table~\ref{tab:efficiency}.

\begin{table}[h]
\centering
\begin{tabular}{cccc}
\hline
Subsystem & Participant & No. Elements in AC & WL1 (minutes)\\ \hline
A          & A      & 102            & 505*\\
A          & B      & 98             & 262\\
A          & C      & 110            & 87\\
B          & A      & 231            & 1143*\\
B          & B      & 227             & 502\\
B          & C      & 246             & 105\\ \hline
\end{tabular}
\caption{Comparative experiment for efficiency evaluation.}
\label{tab:efficiency}
\end{table}

In the experiment for Subsystem A, Participant A required approximately 505 minutes, and produced an assurance case with 102 elements; Participant B required approximately 262 minutes, and produced an assurance case with 98 elements; Participant C required approximately 87 minutes, and produced an assurance case with 110 elements.
It is to be noted that the difference in number of elements in the assurance case is because that creating an assurance case is highly subjective, hence the difference in numbers were expected.
It is also to be noted that since Participant A took a complete manual process, the assurance case produced was NOT a model-based one, therefore it could NOT be converted to runtime assurance case and could NOT be evaluated against runtime data. 
Therefore, the time taken for Participant A shown in table~\ref{tab:efficiency} only reflected time taken following ACCESS steps 1 to 5.

The time breakdown for participants to engineer Subsystem A are: 
\begin{itemize}
    \item Participant A: 45 mins for Step 1; 50 mins for Step 2; 30 mins for Step 3; 180 mins for Step 4; 200 min for Step 5; N/A for Step 6 and 7.
    \item Participant B: 47 mins for Step 1; 20 mins for Step 2; 19 mins for Step 3; 43 mins for Step 4; 56 min for Step 5; 30 min for Step 6; and 47 mins for Step 7.
    \item Participant C: 40 mins for Step 1; 20 mins for Step 2; 19 mins for Step 3; 1 mins for Step 4; 3 min for Step 5; 2 min for Step 6; and 2 mins for Step 7.
\end{itemize}

We also observed that the time it took to follow ACCESS Step 1 for all participants are in the same order of magnitude, this was due to the fact that Participants B and C chose to create modelling languages for their system definition, system requirements and safety concept. 
Although it took them about the same time as Participant A, we argue that the creation of modelling languages is a one-off effort -- in subsequent experiments, the modelling languages are re-used, further reducing the time taken. 

We do not discuss the time taken to engineer Subsystem B in detail, except that Participant B and C took significantly less time in ACCESS Steps 1 and 2 due to the reuse of modelling languages they created in engineering Subsystem A.
We could therefore draw the conclusion that by adopting ACCESS with model-based support, it improves development efficiency comparing to manual effort, even more so when ACME support is available.

\subsection{RQ2: Generality}
\label{sec:generality}
As previously mentioned, we used ACCESS to develop the AUV and a safety-critical autonomous robotic system. 
The model-based tool support for ACCESS (either it being ACME or other model-based assurance case management environment) is underpinned by two fundamental aspects of Model Driven Engineering: 
\begin{itemize}
    \item Domain Specific Modelling, which allows experts in different domains to create modelling languages that best describe their applications;
    \item Model Management Operations, which enable model transformations and model validation in an automated manner.
\end{itemize}

In addition, throughout the development process using ACCESS with model-based support, we found that it was often necessary to ``link'' models defined using different modeling technologies. In our work, this is achieved by exploiting the facilities provided by the Structured Assurance Case Metamodel (SACM). However, other model-based approaches, for example, the use of a \textit{weaving model}~\cite{hawkins2015weaving} may also be adopted to realise the links among heterogeneous models.

In addition, we would like to point out that, some components of System B were developed in conformance to ISO 26262 as SEooCs (Safety Element out of Context)~\cite{iso26262} following the development process defined in ISO 26262. From our experience the ACCCESS methodology integrated seamlessly with the development of SEooCs which would be certified against ISO 26262.

Hence, we draw the conclusion that the ACCESS methodology is a generic approach across different domains in the context of safety critical systems, as long as an assurance case is required for the certification of the target system.

\subsection{RQ3: Coverage}
\label{sec:coverage}
During the development of systems following ACCESS, we needed to deal with models defined using different tools and technologies. 
The types of models included: 
\begin{itemize}
    \item Model defined using Eclipse Modeling Framework (EMF), EMF models are most commonly seen as EMF is the de-facto modelling framework for most open source tools, and are supported by most open source modelling platforms. We also used EMF to model our system requirements, safety concepts, system architectures, etc.;
    \item Simulink models, Simulink is a modelling environment under MATLAB, it provides a graphical block-based modelling framework that supports the design, simulation and analysis of systems; we made use of Simulink models to design some of the schematics of our EE systems;
    \item Excel spreadsheets, spreadsheets are typically used to store Failure Mode and Effect Analysis (FMEA), which is an inductive safety analysis method used in identifying failure modes of components and their effects in the system;
    \item UML models, standardised by the Object Management Group, most UML tools (commercial or open-source) uses XMI (XML Metadata Interchange) format to store models;
    \item Formal models, in our work, Isabelle models are used to verify software behaviour;
    \item JSON models, structured models with no metadata for rapid modelling.
\end{itemize}
In our work, due to the fact that we adopted Eclipse Epsilon as the supporting model management framework, we are able to support the above model formats by either using Epsilon's Model Connectivity (EMC) layer, or directly using Epsilon's existing model driver. 
If there are new types of models defined using new modeling technology, we may also support it by extending the EMC by creating a dedicated model driver for it.
We therefore draw the conclusion that ACCESS supported by ACME achieve a high degree of coverage in managing heterogeneous models.

\subsection{RQ4: Scalability}
\label{sec:scalability}
Our last evaluation is on the scalability of ACME, although ACME is the secondary contribution of our work, we report our findings in the scalability of the tools.
Our evaluation was performed on the premise that the majority of the models used in our development process are EMF models, with a mixture of Simulink models, Excel spreadsheets, formal models and JSON models, which are identified in Section~\ref{sec:generality}. 
To evaluate the scalability of ACME, we selected 5 data sets as shown in Table~\ref{tab:scalability}.
It is to be noted that the in our end result systems, the maximum number of model elements we have in our collection of models was 5689 (Set3). 
We made duplicates of our models and put them together to form Set4 and Set5 to evaluate the scalability of ACME in different order of magnitudes. 
We found that ACME suffered from scalability issues from Set4 and would not load Set5 due to memory overflow. 
This is typically caused by the fact that ACME need to load EMF models in their entirety before any queries can be performed on them, which is an existing issue discovered in various studies~\cite{barmpis2013hawk,wei2016partial,shah2014framework}.

\begin{table}[h]
\centering
\begin{tabular}{ccc}
\hline
Model & No. of Model Elements & Time taken for Evaluation(sec) \\ \hline
Set0          & 109      & 0.1               \\
Set1          & 269      & 0.2                  \\
Set2          & 1369      & 0.8                 \\
Set3          & 5689      & 4.1                 \\
Set4          & 5689000      & 48.3                  \\ \hline
Set5          & 568990000      & N/A                  \\ \hline
\end{tabular}
\caption{Normalised efficiency experiment.}
\label{tab:scalability}
\end{table}

%% file: section_7.tex
\section{Related Work}
\label{sec:related}
There are a number of assurance case works and tools that promote automation by adopting MDE, such as AdvoCATE \cite{denney2017tool}, D-Case Editor \cite{matsuno2010dependability}, ASCE \cite{netkachova2014tool}, Astah GSN \cite{larrucea2017supporting}, and CertWare \cite{barry2011certware}.

MDE is applied from different perspectives of Assurance Case process including
\begin{enumerate}
    \item to generate AC following the process of ``predefined pattern'' and ``pattern instantiation''. The pattern can be modelled by extending the syntax of graphical notations, and the relationship between system data and AC elements can be modelled for automatic instantiation.
    \item to verify by formal verification for evidence generation.
    \item to check the correctness of AC structure by structure modelling.
    \end{enumerate}

NASA has developed a powerful graphical tool AdvoCATE \cite{denney2017tool} based on Eclipse EMF for AC generation, management, and evaluation. MDE is applied in two aspects; GSN is extended with a formal syntax to support the syntactic checks; and the evidence is generated by exploiting FM; it also uses a formal foundation for a lightweight semantic checking based on the metadata attached to the nodes though it is not to encode the whole argument in a formal machine-checkable language. The work follows the process of pattern design, instantiation, and formal verification. The AC pattern designed is generic without specific application constraints to the systems, and is split into two levels, from hazard mitigation to safety requirements, and from safety requirements down to the evidence. The claims at the second level, i.e. the safety requirements, are formalized manually, and verified by invoking the AUTOCERT tool \cite{Denney2008}. The instantiation requires engineers to identify the logic relationship among system data and between AC nodes and system data. The mapping is represented as a table to facilitate the automatic instantiation.

AUTOCERT invoked by AdvoCATE further invokes an automatic theorem prover to verify that the code satisfies the safety properties. However, no means for evaluating the referenced engineering artifacts has been provided.
During operation, the change of the system design and the invalidation of assumptions, etc. can be identified, then ACs are automatically instantiated with updated system data for evolution. However, the AC update process will be partially manual if the safety claim is changed which requires the manual formalization.

Hawkins et al. proposed a model-based approach for generating Assurance Cases \cite{hawkins2015weaving}. Model-based GSN assurance case patterns are used as a basis. 
The work exploits the concept of weaving models \cite{del2006weaving} that represent the links between metamodels.  With the weaving models, the approach allows the instantiation of GSN patterns, which automatically instantiate \textit{weaves} (by means of model transformation) system information into an assurance case based on the links in the weaving models. 
The links between assurance case elements and system data can be updated automatically when the system design changes because the links are built between the metamodels instead of specific system data.  However, apart from injecting system information into an assurance case, no traceability support from an assurance case to engineering artifacts is provided. Consequently, the validation and verification of assurance case is not covered.

\cite{lin2016framework} is also based on a pre-defined pattern and pre-organised data. The software development process is modelled. The process activities' results are software artifacts whose relationships are shown through the input and output of processes. For example,the contribution of software to the hazard derives the software safety requirements. Thus, the relationships of different classes of data can be extracted from the development process models. The pattern is designed specifically for software safety with consideration of software contribution to the hazard. The work adds the syntax of GSN
to allow automatic instantiation of the software artefacts. With the help of a relationship model, the impact of artefact change, e.g. the change of the software contribution to a hazard, can be identified automatically in a AC structure for a convenient review. 
However, since the software is not required to be developed with MDE, the automatic verification of AC is not considered in this work. 

Utsunomiya et al. \cite{utsunomiya2018tool} developed a tool for constructing assurance cases by reading architecture models, quality properties, and risk measurements written in Extensible Markup Language (XML). Assurance cases are generated which conform to SACM v1.0 XML Metadata Interchange schema definition. The tool does not cover the evidence generation of assurance cases. The effectiveness of the method has been insufficiently evaluated to show whether it can be generalised.

Prokhorova et al. \cite{prokhorova2015facilitating} provide a solution for formalizing sub-claims and verifying with formal method tools. The system properties represented by sub-claims are categorised into eight classes such as temporal properties, timing properties, etc. Patterns are established for each class with different formal method verification solutions. The combination of verification tools shows the necessity of multiple formal methods, also referred to as integrated formal methods~\cite{paige1997meta}. 

Gleirscher et al. \cite{gleirscher2019evolution} proposed to model assurance cases formally for autonomous robot systems following the pattern and instantiation process. The purpose is to cope with the AC evolution during the system development instead of the operation. Two assurance case patterns are designed, that cover the construction pattern for the system specification phase, and the extension pattern for the system implementation/ refinement phase. 
Both are instantiated with the system models.
Since the system is required to be modelled in a formal language, there is no need for a process of formalisation of system models, and formal method is a natural choice for assurance case verification. However, the two-phase patterns are specific to the assumption/guarantee (A/G)-style reasoning, therefore not suitable for other RAS systems that do not follow A/G-style reasoning. 

Calinescu et al. \cite{calinescu2017engineering} proposes the dynamic verification of self-adaptive systems at runtime, but no tool support has been provided. The work proposed the assurance case pattern for self-adaptive systems which is instantiated with formalised system requirements, and further verified by model checkers, which facilitates the automatic co-evolution of system and assurance case by avoiding the manual process of system model formalization.

Gacek et al. \cite{gacek2014resolute} proposed a new domain-specific language, Resolute, which is also a tool for building assurance cases based on AADL\cite{as55062004architecture} models.
The generation of assurance cases consists of two steps. 
Firstly, the top-level claim is defined formally by the engineers in first-order predicates where AADL models are queried.
Secondly, the engineers decompose the top-level claim quantitatively. An example could be that the sum of message delay of threads is bounded while the top-level claim is the message delay of the process is bounded. Then the AADL model will be queried through the assurance case and checked automatically by model checking.
The claims and rules for two steps are recorded in the Resolute library for reuse. 
The way to integrate Resolute assurance case with AADL model is to add a “prove" statement within AADL models. 
As the assurance case is written in the same developing environment of AADL models, system model to support assurance case claims will be queried by Resolute models, not transformed or instantiated. So the Resolute assurance case is directly integrated into the system architecture model. This integration enables the traceability and consistency between the system model and its assurance cases, and facilitates the automatic co-evolution of system and assurance case.But this integration on the other hand limits the solution application to other system modelling languages. Since ACME is based on the EMF and the Epsilon framework, it is possible to support various modelling languages (which is demonstrated in Section~\ref{sec:case}).
It is noted that Resolute can only be applied to the system architecture level, rather than the implementation, otherwise it will incur state explosion. Also, the claim of the use case is security properties, where the architectural data are sufficient as assurance case inputs. However, it will require more data (e.g. hazard models, safety-related function models) for safety property argumentation; and Resolute may not be capable of describing those properties properly. 

Rushby conceived of an evidential tool bus~\cite{Rushby2005-ETB} that would allow integration of various verification tools to provide evidence to an assurance case, an idea that was later realised by Cruanes et al.~\cite{Cruanes2013}. Isabelle is also an evidential tool bus, and its connection to ACME allows linking to formal evidence.

%% file: section_8.tex
\section{Summary and Future Work}
\label{sec:conclusion}
In this work, we present ACCESS, a development methodology that promotes the development of safety-critical systems around an evolving model-based assurance case. 
We present ACME, along with ACCESS, and illustrate how ACCESS steps can be followed to develop a critical system from the beginning.

ACME is an integrated model-based assurance case development tool, that supports fine-grained traceability to engineering artifacts such as EMF models, Excel spreadsheets and Isabelle theory files. 
For model-based engineering artifacts, ACME provides the support that enables the users to attach model queries to SACM elements, which can be automatically executed to validate the engineering artifacts.
For external Isabelle theory files, we provide support which makes use of an Isabelle server to process the theory files and reflect the result back to ACME.
In this case, assurance cases can be automatically evaluated, which includes the traceability and validation of engineering artifacts, that significantly reduces development time and improves development efficiency, comparing to human efforts.
In addition, assurance cases can co-evolve with system development, for changes in the system engineering artifacts can be validated by ACME and problems can be found rapidly, to allow a rapid turn-around between system development and system assurance.
We also showed how a development time assurance case can be converted to a dynamic assurance case, with traceability to runtime data. 
With the help of \textit{Dynamic Safety Management System} and \textit{Runtime Data Drivers}, we are able to monitor the validity of the dynamic assurance case in a real time manner.

It is to be noted that we could not show activities and outcomes in each ACCESS step entirely in this paper, due to space limitation.
As such, in the future we would continuously improve the ACCESS methodology and perhaps publish guidelines based on ACCESS to explain the assurance case centric development process in greater details.

With regard to ACME, we have shown that we can trace engineering artifacts such as EMF models, Excel spreadsheets and Isabelle theory files. 
In the future, we plan to support models defined in other modelling technologies such as Simulink and PTC Integrity Modeller. 
Currently, ACME only support EOL for model validaiton rules, which practitioners may not be acquiented with. 
In the future, we plan to lower the technical barrier by providing support to \textit{Constrained Natural Language} as query language, so that it requires minimal effort to learn.
Also, the impact of system engineering artifacts to assurance case currently requires the practitioners to evaluate the assurance case to detect invalidities. 
In the future we plan to provide support for passive impact analysis with a version control system, so that we can support version control on both assurance cases and engineering artifacts.
Finally, we plan to support SACM's argument notation alongside with GSN notation, to prompt the wider adoption of SACM.

%% file: main.bbl
\begin{thebibliography}{10}

\bibitem{Alkassar2008}
E.~Alkassar, M.~Hillebrand, D.~Leinenbach, N.~Schirmer, and A.~Starostin.
\newblock The {Verisoft} approach to systems verification.
\newblock In {\em VSTTE 2008}, volume 5295 of {\em LNCS}, pages 209--224. Springer, 2008.

\bibitem{as55062004architecture}
S.~AS5506.
\newblock Architecture analysis and design language (aadl).
\newblock {\em Embedded Computing Systems Committee, SAE}, 2004.

\bibitem{barmpis2013hawk}
K.~Barmpis and D.~Kolovos.
\newblock Hawk: Towards a scalable model indexing architecture.
\newblock In {\em Proceedings of the Workshop on Scalability in Model Driven Engineering}, pages 1--9, 2013.

\bibitem{barry2011certware}
M.~R. Barry.
\newblock Certware: A workbench for safety case production and analysis.
\newblock In {\em Aerospace Conference, 2011 IEEE}, pages 1--10. IEEE, 2011.

\bibitem{bishop2000methodology}
P.~Bishop and R.~Bloomfield.
\newblock A methodology for safety case development.
\newblock In {\em Safety and Reliability}, volume~20, pages 34--42. Taylor \& Francis, 2000.

\bibitem{Blanchette2011}
J.~C. Blanchette, L.~Bulwahn, and T.~Nipkow.
\newblock Automatic proof and disproof in {Isabelle/HOL}.
\newblock In {\em FroCoS}, volume 6989 of {\em LNCS}, pages 12--27. Springer, 2011.

\bibitem{brambilla2017model}
M.~Brambilla, J.~Cabot, and M.~Wimmer.
\newblock Model-driven software engineering in practice.
\newblock {\em Synthesis lectures on software engineering}, 3(1):1--207, 2017.

\bibitem{Brookes1984}
S.~D. Brookes, C.~A.~R. Hoare, and A.~W. Roscoe.
\newblock A theory of communicating sequential processes.
\newblock {\em Journal of the ACM}, 31(3):560--599, 1984.

\bibitem{Brucker2019-DOFCert}
A.~Brucker and B.~Wolff.
\newblock Using ontologies in formal developments targeting certification.
\newblock In {\em Integrated Formal Methods (iFM)}, volume 11918 of {\em LNCS}, pages 65--82. Springer, 2019.

\bibitem{calinescu2017engineering}
R.~Calinescu, D.~Weyns, S.~Gerasimou, M.~U. Iftikhar, I.~Habli, and T.~Kelly.
\newblock Engineering trustworthy self-adaptive software with dynamic assurance cases.
\newblock {\em IEEE Transactions on Software Engineering}, 44(11):1039--1069, 2018.

\bibitem{Cruanes2013}
S.~Cruanes, G.~Hamon, S.~Owre, and N.~Shankar.
\newblock Tool integration with the evidential tool bus.
\newblock In {\em VMCAI}, volume 7737 of {\em LNCS}, pages 275--294. Springer, 2013.

\bibitem{del2006weaving}
M.~D. Del~Fabro, J.~B{\'e}zivin, and P.~Valduriez.
\newblock Weaving models with the eclipse amw plugin.
\newblock In {\em Eclipse Modeling Symposium, Eclipse Summit Europe}, volume 2006, pages 37--44, 2006.

\bibitem{denney2017tool}
E.~Denney and G.~Pai.
\newblock Tool support for assurance case development.
\newblock {\em Automated Software Engineering}, pages 1--65, 2017.

\bibitem{denney2015dynamic}
E.~Denney, G.~Pai, and I.~Habli.
\newblock Dynamic safety cases for through-life safety assurance.
\newblock In {\em 2015 IEEE/ACM 37th IEEE International Conference on Software Engineering}, volume~2, pages 587--590. IEEE, 2015.

\bibitem{Denney2008}
E.~Denney and S.~Trac.
\newblock {A software safety certification tool for automatically generated guidance, navigation and control code}.
\newblock {\em IEEE Aerospace Conference Proceedings}, 2008.

\bibitem{gmf}
{Eclipse Foundation}.
\newblock {Eclipse Modelling Framework (GMF)}.
\newblock https://www.eclipse.org/modeling/gmp/.

\bibitem{eurocontrol}
{European Organisation for the Safety of Air Navigation (EUROCONTROL)}.
\newblock {\em Safety Case Development Manual}.
\newblock 2006.

\bibitem{Foster2018-FACS}
S.~Foster, J.~Baxter, A.~Cavalcanti, A.~Miyazawa, and J.~Woodcock.
\newblock Automating verification of state machines with reactive designs and {Isabelle/UTP}.
\newblock In {\em Proc. 15th. Intl. Conf. on Formal Aspects of Component Software}, volume 11222 of {\em LNCS}. Springer, October 2018.

\bibitem{Foster2019-iFM}
S.~Foster, Y.~Nemouchi, M.~Gleirscher, and T.~Kelly.
\newblock {Isabelle/SACM}: Computer-assisted assurance cases with integrated formal methods.
\newblock In {\em iFM}, LNCS 11918, pages 379--398. Springer, December 2019.

\bibitem{foster2020formal}
S.~D. Foster, Y.~Nemouchi, C.~O'Halloran, N.~Tudor, and K.~Stephenson.
\newblock Formal model-based assurance cases in isabelle/sacm: An autonomous underwater vehicle case study.
\newblock In {\em Formal Methods in Software Engineering (FormaliSE 2020): Proceedings of the 8th International Conference}. ACM, 2020.

\bibitem{gacek2014resolute}
A.~Gacek, J.~Backes, D.~Cofer, K.~Slind, and M.~Whalen.
\newblock Resolute: an assurance case language for architecture models.
\newblock {\em ACM SIGAda Ada Letters}, 34(3):19--28, 2014.

\bibitem{gleirscher2019evolution}
M.~Gleirscher, S.~Foster, and Y.~Nemouchi.
\newblock Evolution of formal model-based assurance cases for autonomous robots.
\newblock In {\em International Conference on Software Engineering and Formal Methods}, pages 87--104. Springer, 2019.

\bibitem{Gleirscher2018-NewOpportunitiesIntegrated}
M.~Gleirscher, S.~Foster, and J.~Woodcock.
\newblock New opportunities for integrated formal methods.
\newblock {\em ACM Computing Surveys}, 52(6), 2019.

\bibitem{greenwell2006taxonomy}
W.~S. Greenwell, J.~C. Knight, C.~M. Holloway, and J.~J. Pease.
\newblock A taxonomy of fallacies in system safety arguments.
\newblock 2006.

\bibitem{habli2018safety}
I.~Habli, S.~White, M.~Sujan, S.~Harrison, and M.~Ugarte.
\newblock What is the safety case for health it? a study of assurance practices in england.
\newblock {\em Safety Science}, 110:324--335, 2018.

\bibitem{hawkins2015weaving}
R.~Hawkins, I.~Habli, D.~Kolovos, R.~Paige, and T.~Kelly.
\newblock Weaving an assurance case from design: a model-based approach.
\newblock In {\em High Assurance Systems Engineering (HASE), 2015 IEEE 16th International Symposium on}, pages 110--117. IEEE, 2015.

\bibitem{iaea}
{International Atomic Energy Agency (IAEA)}.
\newblock {\em IAEA Safety Glossary: Terminology Used in Nuclear Safety and Radiation Protection}.
\newblock 2008.

\bibitem{iso26262}
{International Organization for Standardization (ISO)}.
\newblock {\em ISO 26262: Road Vehicles - Functional Safety.}
\newblock 2011.

\bibitem{jaaksi2002developing}
A.~Jaaksi.
\newblock {Developing Mobile Browsers in a Product Line}.
\newblock {\em IEEE software}, 19(4):73--80, 2002.

\bibitem{karna2009evaluating}
J.~K{\"a}rn{\"a}, J.-P. Tolvanen, and S.~Kelly.
\newblock {Evaluating the Use of Domain-Specific Modeling in Practice}.
\newblock In {\em {Proceedings of the 9th OOPSLA workshop on Domain-Specific Modeling}}, 2009.

\bibitem{kelly2004goal}
T.~Kelly and R.~Weaver.
\newblock The goal structuring notation--a safety argument notation.
\newblock In {\em Proceedings of the dependable systems and networks 2004 workshop on assurance cases}, page~6. Citeseer, 2004.

\bibitem{kelly1999arguing}
T.~P. Kelly.
\newblock {\em Arguing safety: a systematic approach to managing safety cases}.
\newblock PhD thesis, University of York York, UK, 1999.

\bibitem{kolovos2006epsilon}
D.~S. Kolovos, R.~F. Paige, and F.~A. Polack.
\newblock The epsilon object language (eol).
\newblock In {\em European Conference on Model Driven Architecture-Foundations and Applications}, pages 128--142. Springer, 2006.

\bibitem{kolovos2008epsilon}
D.~S. Kolovos, R.~F. Paige, and F.~A. Polack.
\newblock The epsilon transformation language.
\newblock In {\em International Conference on Theory and Practice of Model Transformations}, pages 46--60. Springer, 2008.

\bibitem{larrucea2017supporting}
X.~Larrucea, A.~Walker, and R.~Colomo-Palacios.
\newblock Supporting the management of reusable automotive software.
\newblock {\em IEEE Software}, (3):40--47, 2017.

\bibitem{Lee2018}
E.~A. Lee and M.~Sirjani.
\newblock What good are models?
\newblock In {\em FACS}, volume 11222 of {\em LNCS}. Springer, 2018.

\bibitem{lin2016framework}
C.-L. Lin, W.~Shen, and S.~Drager.
\newblock A framework to support generation and maintenance of an assurance case.
\newblock In {\em 2016 IEEE International Symposium on Software Reliability Engineering Workshops (ISSREW)}, pages 21--24. IEEE, 2016.

\bibitem{Machin2018SMOF}
M.~Machin, J.~Guiochet, H.~Waeselynck, J.-P. Blanquart, M.~Roy, and L.~Masson.
\newblock {SMOF}: A safety monitoring framework for autonomous systems.
\newblock {\em IEEE Transactions on Systems, Man, and Cybernetics}, 48(5), May 2018.

\bibitem{simulink}
{Mathworks}.
\newblock {Simulink}.
\newblock \url{https://www.mathworks.com/products/simulink.html}.
\newblock Online; accessed 6th June, 2020.

\bibitem{matsuno2010dependability}
Y.~Matsuno, H.~Takamura, and Y.~Ishikawa.
\newblock A dependability case editor with pattern library.
\newblock In {\em High-Assurance Systems Engineering (HASE), 2010 IEEE 12th International Symposium on}, pages 170--171. IEEE, 2010.

\bibitem{mcdermid2001software}
J.~A. McDermid.
\newblock Software safety: where's the evidence?
\newblock In {\em Proceedings of the Sixth Australian workshop on Safety critical systems and software-Volume 3}, pages 1--6. Australian Computer Society, Inc., 2001.

\bibitem{miyazawa2016robochart}
A.~Miyazawa, P.~Ribeiro, W.~Li, A.~Cavalcanti, J.~Timmis, and J.~Woodcock.
\newblock Robochart: a state-machine notation for modelling and verification of mobile and autonomous robots.
\newblock {\em Tech. Rep.}, 2016.

\bibitem{Miyazawa2019-RoboChart}
A.~Miyazawa, P.~Ribeiro, W.~Li, A.~Cavalcanti, J.~Timmis, and J.~Woodcock.
\newblock Robochart: modelling and verification of the functional behaviour of robotic applications.
\newblock {\em Software and Systems Modelling}, January 2019.

\bibitem{nair2015evidence}
S.~Nair, J.~L. de~la Vara, M.~Sabetzadeh, and D.~Falessi.
\newblock Evidence management for compliance of critical systems with safety standards: A survey on the state of practice.
\newblock {\em Information and Software Technology}, 60:1--15, 2015.

\bibitem{netkachova2014tool}
K.~Netkachova, O.~Netkachov, and R.~Bloomfield.
\newblock Tool support for assurance case building blocks.
\newblock In {\em International Conference on Computer Safety, Reliability, and Security}, pages 62--71. Springer, 2014.

\bibitem{Nipkow2014-ConcreteSemantics}
T.~Nipkow and G.~Klein.
\newblock {\em Concrete Semantics with {Isabelle/HOL}}.
\newblock Springer, December 2014.

\bibitem{Isabelle}
T.~Nipkow, M.~Wenzel, and L.~C. Paulson.
\newblock {\em {Isabelle/HOL: A Proof Assistant for Higher-Order Logic}}, volume 2283 of {\em LNCS}.
\newblock Springer, 2002.

\bibitem{sacm}
{Object Management Group}.
\newblock {Structured Assurance Case Metamodel}.
\newblock \url{https://www.omg.org/spec/SACM}.
\newblock Online; accessed 6th June, 2020.

\bibitem{paige1997meta}
R.~F. Paige.
\newblock A meta-method for formal method integration.
\newblock In {\em International Symposium of Formal Methods Europe}, pages 473--494. Springer, 1997.

\bibitem{prokhorova2015facilitating}
Y.~Prokhorova, L.~Laibinis, and E.~Troubitsyna.
\newblock Facilitating construction of safety cases from formal models in event-b.
\newblock {\em Information and Software Technology}, 60:51--76, 2015.

\bibitem{rose2008epsilon}
L.~M. Rose, R.~F. Paige, D.~S. Kolovos, and F.~A. Polack.
\newblock The epsilon generation language.
\newblock In {\em European Conference on Model Driven Architecture-Foundations and Applications}, pages 1--16. Springer, 2008.

\bibitem{Rushby2005-ETB}
J.~Rushby.
\newblock An evidential tool bus.
\newblock In {\em Formal Methods and Software Engineering (ICFEM)}, volume 3785 of {\em LNCS}. Springer, 2005.

\bibitem{shah2014framework}
S.~M. Shah, R.~Wei, D.~S. Kolovos, L.~M. Rose, R.~F. Paige, and K.~Barmpis.
\newblock A framework to benchmark nosql data stores for large-scale model persistence.
\newblock In {\em International Conference on Model Driven Engineering Languages and Systems}, pages 586--601. Springer, 2014.

\bibitem{steinberg2008emf}
D.~Steinberg, F.~Budinsky, E.~Merks, and M.~Paternostro.
\newblock {\em EMF: eclipse modeling framework}.
\newblock Pearson Education, 2008.

\bibitem{trapp2013safety}
M.~Trapp, D.~Schneider, and P.~Liggesmeyer.
\newblock A safety roadmap to cyber-physical systems.
\newblock In {\em Perspectives on the future of software engineering}, pages 81--94. Springer, 2013.

\bibitem{Tuong2019-CIsabelle}
F.~Tuong and B.~Wolff.
\newblock Deeply integrating {C11} code support into {Isabelle/PIDE}.
\newblock In {\em Formal Integrated Development Environment (F-IDE)}, volume 310 of {\em EPTCS}, pages 13--28, 2019.

\bibitem{jsp430}
{U.K. Ministry of Defence (MOD)}.
\newblock {\em JSP 430 - Ship Safety Management System Handbook}.
\newblock 1996.

\bibitem{0055}
{U.K. Ministry of Defence (MOD)}.
\newblock {\em 00-55 Requirements of Safety Related Software in Defence Equipment}.
\newblock 1997.

\bibitem{ukmod}
{U.K. Ministry of Defence (MOD)}.
\newblock {\em Safety Management Requirements for Defence Systems}.
\newblock 2007.

\bibitem{ukrail}
{U.K. Rail Safety Standards Board}.
\newblock {\em Engineering Safety Management Issue 4}.
\newblock 2007.

\bibitem{utsunomiya2018tool}
H.~Utsunomiya, N.~Kobayashi, S.~Morisaki, and S.~Yamamoto.
\newblock A tool to create assurance case through models.
\newblock {\em Transactions on Machine Learning and Artificial Intelligence}, 6(2):46--46, 2018.

\bibitem{wei2022designing}
R.~Wei, Z.~Jiang, X.~Guo, H.~Mei, A.~Zolotas, and T.~Kelly.
\newblock Designing critical systems with iterative automated safety analysis.
\newblock In {\em Proceedings of the 59th ACM/IEEE Design Automation Conference}, pages 181--186, 2022.

\bibitem{wei2023decisive}
R.~Wei, Z.~Jiang, X.~Guo, R.~Yang, H.~Mei, A.~Zolotas, and T.~Kelly.
\newblock Decisive: Designing critical systems with iterative automated safety analysis.
\newblock {\em IEEE Transactions on Computer-Aided Design of Integrated Circuits and Systems}, 2023.

\bibitem{wei2023automated}
R.~Wei, Z.~Jiang, H.~Mei, K.~Barmpis, S.~Foster, T.~Kelly, and Y.~Zhuang.
\newblock Automated model based assurance case management using constrained natural language.
\newblock {\em IEEE Transactions on Computer-Aided Design of Integrated Circuits and Systems}, 2023.

\bibitem{wei2019model}
R.~Wei, T.~P. Kelly, X.~Dai, S.~Zhao, and R.~Hawkins.
\newblock Model based system assurance using the structured assurance case metamodel.
\newblock {\em Journal of Systems and Software}, 154:211--233, 2019.

\bibitem{wei2016partial}
R.~Wei, D.~S. Kolovos, A.~Garcia-Dominguez, K.~Barmpis, and R.~F. Paige.
\newblock Partial loading of xmi models.
\newblock In {\em Proceedings of the ACM/IEEE 19th International Conference on Model Driven Engineering Languages and Systems}, pages 329--339, 2016.

\bibitem{wei2018}
R.~Wei, J.~Reich, T.~Kelly, and S.~Gerasimou.
\newblock On the transition from design time to runtime model-based assurance cases.
\newblock In {\em 13th International Workshop on Models@Runtime, ACM/IEEE 21th International Conference on Model Driven Engineering Languages and Systems (MoDELS 2018)}, 2018.

\bibitem{Wenzel2019-Isar}
M.~Wenzel.
\newblock Interaction with formal mathematical documents in {Isabelle/PIDE}.
\newblock In {\em CICM}, LNCS 11617, pages 1--15. Springer, 2019.

\bibitem{Wenzel2007FMIsabelle}
M.~Wenzel and B.~Wolff.
\newblock Building formal method tools in the {Isabelle/Isar} framework.
\newblock In {\em TPHOLs}, volume 4732 of {\em LNCS}. Springer, 2007.

\end{thebibliography}
